\documentclass[twocolumn]{pasj00}

\SetRunningHead{T. Miyakawa et al.}{ }
\title{A Variable Partial Covering Model for the Seyfert 1 Galaxy MCG-6-30-15}
\author{Takehiro \textsc{Miyakawa}\altaffilmark{1,2}, Ken
\textsc{Ebisawa}\altaffilmark{1,2}, and Hajime
\textsc{Inoue}\altaffilmark{1}} 
\altaffiltext{1}{The Institute of Space and Astronautical Science (ISAS), Japan
Aerospace Exporlation Agency (JAXA), \\ 3-1-1 Yoshinodai, Chuo, Sagamihara, Kanagawa 252-5210}
\altaffiltext{2}{Department of Astronomy, Graduate School of Science,
the University of Tokyo,\\
7-3-1 Hongo, Bunkyo-ku, Tokyo 113-0033}

\email{miyakawa.takehiro@jaxa.jp}
\Received{$\langle$2010-05-29 $\rangle$}
\Accepted{$\langle$2010 $\rangle$}
\Published{$\langle$publication date$\rangle$}
\begin{document}
\maketitle

\begin{abstract}
We propose a simple spectral model for the Seyfert 1 Galaxy MCG-6-30-15
that can explain   most of 
the 1 -- 40 keV spectral variation by change of the partial covering fraction,
similar to the one  proposed by Miller et al.\ (2008).
Our  spectral model  is composed of three continuum components;
(1) a direct power-law component, (2) 
a heavily absorbed power-law component by mildly ionized intervening matter,  and 
(3) a cold disk reflection component far from the black hole with   moderate 
 solid-angle ($\Omega/2\pi \approx 0.3$) accompanying  a narrow fluorescent iron line. 
The first two components are affected by the surrounding highly ionized thin absorber with 
$N_H\approx 10^{23.4}$cm$^{-2}$ and $\log \xi \approx 3.4$.
The heavy absorber in the second component is fragmented into many clouds, each of which is
composed of radial zones with different ionization states and column densities, the 
 main body ($N_H\approx10^{24.2}$cm$^{-2}$,
$\log \xi \approx 1.6$), the envelope ($N_H\approx 10^{22.1}$cm$^{-2}$,
$\log \xi \approx 1.9$) and presumably a completely opaque core.
These parameters of the ionized absorbers,  as well as the intrinsic spectral shape
of the X-ray source, 
are unchanged at all.
The central X-ray source is moderately extended, and its luminosity is not significantly 
variable. The  observed  flux and spectral variations are 
mostly explained by variation of the geometrical  partial covering fraction of the central 
source  from 0 (uncovered) to $\sim$0.63 by the intervening ionized clouds in the line of sight.
The ionized iron K-edge of the heavily
	  absorbed  component  explains most of 
 the seemingly broad line-like  
	  feature, a well-known spectral characteristic of MCG-6-30-15.
The direct component and the absorbed component anti-correlate, cancelling
their variations each other, so that the fractional 
spectral variation becomes the minimum at the iron energy band; 
another observational characteristic of MCG-6-30-15 is thus  explained.

\end{abstract}

\section{Introduction}

The Seyfert 1 galaxy MCG-6-30-15 is the primary 
``disk-line'' target, in which  a relativistically broadened iron
K-emission line is suggested  to  originate in  the innermost region
of the accretion disk around the 
black hole (e.g., Tanaka et al.\ 1995).
 MCG-6-30-15 is also known to exhibit characteristic X-ray spectral variations;
Inoue \& Matsumoto (2001, 2003), Fabian et al.\ (2002)  
and Matsumoto et al.\ (2003)  reported
significantly small variability in the  iron line energy band of MCG
6-30-15. 
It is found that the energy-dependent Root Mean Square (RMS) variability is suppressed around the iron 
line energy, particularly in longer timescales.
Inoue and Matsumoto (2001, 2003) 
proposed  that the absorbed spectrum due to photo-ionized
warm absorbers mimic shape of the strongly red-shifted iron line, 
and  variation of the warm absorbers may explain the  apparently
small variability in the iron line energy band. 

On the other hand, 
Miniutti and Fabian (2004) proposed that the suppressed 
variability in the iron line band may be  explained by  the 
general relativistic light-bending effects, which take place
in the very vicinity of the black hole. 
Miniutti et al.\ (2007) claim that the 
Suzaku spectrum of MCG-6-30-15 is in fact consistent with the light-bending model.
Nied\'{z}wiecki \& \.{Z}ycki (2008) 
and 
Nied\'{z}wiecki \& Miyakawa (2010) 
independently re-examined
the light-bending model, and concluded that it is possible  
to explain the suppressed variability in the iron energy band
assuming a particular configuration and 
movement of the  illuminating source.
\.Zycki et al.\ (2010), on the other hand,  claim
 that the light-bending model is not able to explain the Suzaku spectrum, if 
a broader energy range is adopted than used 
by Miniutti et al.\ (2007).


The primary question in MCG-6-30-15 is  whether the iron line is truely 
broad and its little variability is a consequence  of the general relativistic light bending effects
or not.
 The critical point  to address the problem
is to model the underlying continuum spectrum correctly, 
since the broad iron line parameters are  dependent on the choice of 
the continuum spectral model.
In fact, on one hand, Miller, Turner, and Reeves (2008) claim that the 
``disk-line'' is not required, when the continuum spectrum is modeled with
multiple warm absorbers.
On the other hand, 
 Reynolds et al. (2009) argue  that the absorption-dominated model 
over-predicts the  6.4 keV fluorescent line emission, and the disk-line model is more
physically reasonable.
 Miller et al. (2009) counterargue  that the calculation in Reynolds et al. (2009) 
neglects opacity at the 6.4 keV line energy, and conclude  that 
variation in the partial-covering fraction may dominate the observed 
 X-ray spectral variability. However, the 
partial covering scenario by Miller et al. (2009) requires a number of free parameters
to model the observed spectrum.  Therefore, in order to examine the 
partial covering scenario, reducing the number of  parameters, if possible, and
constructing a concrete physical picture to explain the  spectral variation is certainly intriguing.

Recently,  using the Suzaku data taken in January 2006, Miyakawa et
al. (2009) (hereafter Paper I) found a clear correlation between the 
intensity in the 6 -- 10 keV band and the spectral ratio of 0.5 -- 3.0
keV / 6 -- 10 keV in a time-scale of $\lesssim 10^5$ sec.
Essentially the same spectral 
correlation was confirmed for longer timescales up to $\sim$14 years
in the  RXTE data (Miyakawa 2010).
Paper I  adopted a spectral model composed of a 
direct power-law component, its  reflection component by a neutral matter
(with the solid-angle $\Omega/2\pi \approx 1$), two warm absorbers 
with different ionization states. 
The observed 
 spectral variation requires change of the apparent slope of the direct
component, whereas the shape and intensity of the reflection component
 being invariable. 
A mildly broad 
iron emission line was  required at $6.42\pm0.06$ keV
with an intrinsic width of 0.29 keV (1 $\sigma$) and the equivalent-width
100 eV; such an extremely broadened ``disk line'' as is claimed
by Miniutti et al.\ (2007) was not required. 

In this paper, following Paper I,
we attempt to comprehensively understand  spectral 
variability of MCG-6-30-15 using Suzaku and Chandra archival data. 
Our goal is to 
find a   spectral model 
 which can naturally  explain the observed  spectral variation 
of MCG-6-30-15 with a minimum set of free-parameters.
If we are successful to construct such a reasonable
spectral model, we will see if the relativistically distorted iron emission
line  is truely present or not in the energy spectrum.
We believe this is an important step forward  to fully understand
 X-ray spectral properties of MCG-6-30-15,  Seyfert galaxies, and black holes.

\section{Observation and Data Reduction}

We use the same Suzaku data 
used in Paper I, taken in 2006 from  January 9 to 14th  (143 ksec exposure),
from 23 to 26  (99 ksec) and from 27 to 30  (97 ksec).
We use the XIS data in 1 -- 10 keV, and PIN data in 10 -- 40 keV.
In the present paper, we are interested in spectral variations above 1 keV, so 
we did not use the data below 1 keV.
All  of the spectral fits were made with XSPEC v11 (Arnaud 1996). 
In the following, the xspec model names used in the spectral analysis  are explicitly given.
 A constant factor to adjust normalizations between  XIS and PIN  is fixed at 1:1.086 (Ishida et al. 2007). 
For more details about Suzaku data reduction,  see Paper I.


The Chandra satellite observed MCG-6-30-15 several times.
In this paper, we use the 
High Energy Transmission Grating Spectrometer (HETGS) 
data taken between  2004 May 19 and 27, when
the source was observed four times 
resulting in a good exposure time of 522 ksec. 
For data reduction we used the CIAO 4.0 software package; 
``tgextract'' is used to produce  PHA2 spectral  files from
the Level2 event file. 
HETGS is composed of High Energy Grating (HEG) and Medium Energy 
Grating (MEG).
 The $+1$ and $-1$ orders of the HEG were
combined, and so were the $+1$ and $-1$ orders of the MEG. The HEG and
MEG spectra were fitted  separately.  

Errors quoted in this paper are at statistical 90$\%$ confident level.


\section{Data Analysis and Results}
\label{section:model1.specvar} 

\subsection{Introduction of  the ``three-component'' model}

In Paper I, spectral variation of MCG-6-30-15 was studied 
assuming the ``two-component''   spectral model given as,

\begin{equation}
F =  W_H W_L( N_D + N_R R ) P + I_{Fe},
\label{Eq1}
\end{equation}
where $P$ is the intrinsic power-law spectrum with a high-energy cut-off
(fixed at 160 keV; Guainazzi et al. 1999), 
$N_D$ is   normalization factor of the directly observed component,
$N_R$ is  normalization factor of the reflected component,
$R$ is reflection albedo by the cold optically thick matter
(``pexrav'' in xspec: Magdziarz $\&$ Zdziarski 1995),
$W_H$ and $W_L$ represent attenuation by a high- and a low-ionized warm absorbers, 
respectively, 
and
$I_{Fe}$ is  an iron $K_{\alpha}$ emission  line.
Effect of interstellar absorption is always taken into account 
in the spectral fitting 
(``phabs''; Balucinska-Church $\&$ McCammon 1992), 
but its term is omitted  for simplicity
in the above expression 
and hereafter.  Amount of the disk reflection is measured with 
$\Omega$,  solid angle of the reflector seen from the central source. 
Ratio of the normalization of the reflection component to that of  the cutoff power-law component corresponds to 
$\Omega$/2$\pi$,  assuming isotropic emission of the central source.

We used  XSTAR Version 2.1kn8 (Kallman et al. 2004) 
to model the warm absorbers.
The temperature, pressure and density of the warm absorbers are assumed to be  ${10}^5$ K,  0.03 dyne ${\rm cm}^{-2}$ and ${10}^{12} {\rm cm}^{-3}$, respectively.  The incident photo-ionizing spectrum is 
assumed to be a power-law with the index 2.0, and the solar abundance by
Greeves, Noels and Sauval (1996) is adopted for the ionized material.
We made a grid model by running XSTAR for different values of $\xi$ and $N_H$; the log $\xi$ values are from 0.1 to  5 (erg cm s$^{-1}$)  and the $N_H$ values are from ${10}^{20}$ to  ${10}^{24}$ (\rm cm$^{-2}$). The number of
steps for log $\xi$ and $N_H$ are both 20, thus our grid model has $20 \times 20$ grid-points. 
The redshift of a warm absorber and the source  are fixed at 0.001 and 0.00775,
 respectively (Young et al. 2005, Fisher et al. 1995).

In Paper I, we   found that the spectral variation 
at various timescales is successfully explained 
 by variations of  two parameters,   normalization factor of the directly 
observed component ($N_D$) and 
apparent slope of the direct
component, such that the slope gets steeper for greater normalizations. 
 The apparent slope change is described by
either change of the index of the power-law component ($P$), 
ionization state of the 
low-ionized warm absorber, or column density
of the low-ionized warm absorber ($W_L$).

Below, we scrutinize the spectral model in Paper I  carefully, 
and introduce a  more elaborated model.
Using the Chandra HETGS,  Young et
al.\  (2005) resolved a  narrow emission line at 6.4 keV in MCG-6-30-15, which
we  confirmed via re-analysis of  the archival data.
The equivalent width of the line is measured as  $\sim 20$ eV, 
which requires  a cold reflector 
with a solid angle of $\Omega$/2$\pi$ $\sim$ 0.3 viewed from the central X-ray source 
(e.g., George $\&$ Fabian 1991).
On the other hand, the best-fit values of $N_D$ and $N_R$ obtained 
in Paper I indicates that  
$N_R/N_D = \Omega/2\pi \approx 1$.
Consequently, in order to account for  the observed 
 disk reflection spectrum of an amount of $\Omega$/2$\pi$ $\sim$ 1
including the  outer disk reflection with $\Omega$/2$\pi$ $\sim$ 0.3,
we need another disk reflection component or a different spectral component
which is similar to  disk reflection,  corresponding to an amount of
$\Omega$/2$\pi$ $\sim$ 0.7.

Thus, we separate the $N_R$ term in (\ref{Eq1}) into two terms.
One is the reflection continuum component from a cold reflector 
emitting a narrow iron $K_\alpha$ line,
solid angle  of which is $\Omega$/2$\pi$ $\sim$ 0.3.
The other  is introduced to represent the  remaining part of the $N_R$ term;
for this, we adopt a heavily absorbed component by warm absorber, since
the absorbed component is similar to the reflection component in shape,
and does not have to accompany the fluorescent iron emission line.
In fact,  a  partial covering model has been proposed 
to explain the continuum spectral shape and variation of 
MCG-6-30-15  (e.g., Matsuoka et al.\ 1990; McKernan and Yaqoob 1998; 
Miller, Turner and Reeves  2008, 2009).

Resultantly, we introduce a new spectral model  expressed as
\begin{equation}
F = W_H W_L ( N_1 + N_2 W_2 ) P + N_3 R P + I_{Fe} ,
\label{Eq2}
\end{equation}
where $N_1$ is  normalization 
of the direct component, $N_2$ is that for the heavily 
absorbed component,  $W_2$ stands for another warm absorber
that is for the  ionized optically thick partial absorber.
The normalization factor of the outer reflection component, 
$N_3$ is constant and related to the average  of $N_1$ so that 
$N_3/<N_1> = \Omega/2\pi \sim   0.3$, and $I_{Fe}$ is a narrow iron emission
line with a fixed equivalent width of $20$ eV. 
Hereafter, this model is referred as the ``three-component'' model.

\begin{figure}[htbp]
\begin{center}
\FigureFile(90mm,90mm){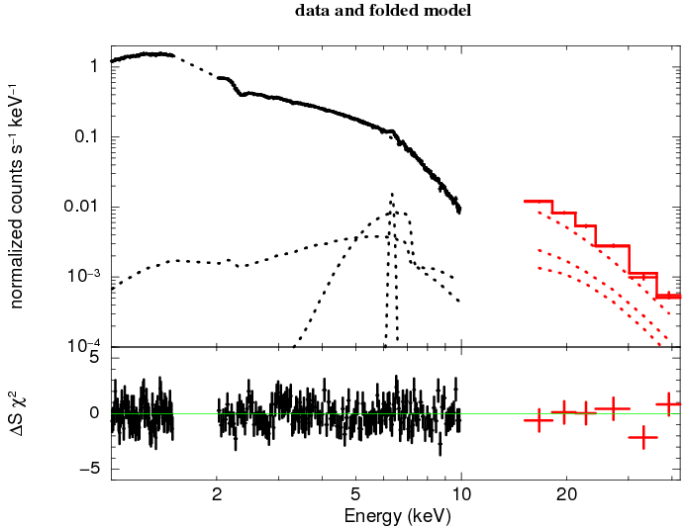}
\end{center}
\caption{Spectral fit result for the 1-40 keV average spectrum  for 
Suzaku XIS/PIN with the three-component model. } 
 \label{Miyakawa_model_fit_Suzaku}
\end{figure}

We try the three-component model to the Suzaku average spectrum in 1 -- 40 keV.
We found the fit improves if we additionally put a weak iron
neutral edge at 7.11 keV and an absorption
line at 7.0 keV.  The former is considered to be  due to a slight difference 
of the assumed $\Omega$ from the exact value,  
and the latter due to  deficiency 
of the H-like iron absorption line equivalent-width already
included in the XSTAR model.
Figure $\ref{Miyakawa_model_fit_Suzaku}$ and  
Table \ref{Miyakawa_model_ave_Suzaku} show the fitting result for
the Suzaku average spectrum with the three-component model in 1 -- 40 keV. 
Parameters of the high-ionized and low-ionized warm absorbers
are similar to those in Paper I.
For the newly introduced  absorption component ($W_2$),
we obtain  $N_H \sim$ 1.6 $\times$ 10$^{24}$ cm$^{-2}$
and  $\log \xi = 1.57$. 
Normalizations of the heavily absorbed component and 
the direct component are  0.60 and 1.44 ($10^{-2}$ photons s$^{-1}$cm$^{-2}$ at 1 keV),  
respectively, which 
suggests that the source is partially covered
by the ionized thick matter by $0.6/(1.44+0.6)\approx$ 30 \%.

We note that the absorbing matter is marginally Compton thick, so that the Compton
scattering may not be negligible for precise spectral calculation  (e.g., Yaqoob 1997).
This is not considered in the present paper, and will be  a subject for future work.

\setlength{\tabcolsep}{2.4pt}
\begin{table}
\begin{center}
\caption{Results of spectral fitting in 1--40 keV for the Suzaku XIS/PIN
average spectrum with the three-component model. }
\begin{tabular}{cc}
\hline\hline
\multicolumn{2}{l}{\em Interstellar absorption}\\
 $N_H$ (${10}^{21}$ ${\rm cm}^{-2}$)  & 1.5$^{+0.4}_{-0.2}$ \\
\hline
\multicolumn{2}{l}{\em $W_H$}\\
$N_H$  (${10}^{23}$ ${\rm cm}^{-2}$) & 2.4$^{+1.8}_{-1.6}$  \\ 
log $\xi$  & 3.37$\pm$0.04 \\ 
\hline
\multicolumn{2}{l}{\em $W_L$}\\
$N_H$  (${10}^{21}$ ${\rm cm}^{-2}$)  & 3.7$\pm$1.3 \\ 
log $\xi$  & 1.54$^{+0.25}_{-0.13}$  \\ 
\hline
\multicolumn{2}{l}{\em $N_1$}\\
10$^{-2}$ ph/s/cm$^2$ at 1 keV &  1.44$\pm$0.03 \\ 
\hline
\multicolumn{2}{l}{\em $N_2$}\\
10$^{-2}$ ph/s/cm$^2$ at 1 keV &  0.60$\pm$0.09 \\
\hline
\multicolumn{2}{l}{\em $W_2$}\\
$N_H$  (${10}^{24}$ ${\rm cm}^{-2}$)  & 1.6$^{+0.6}_{-0.1}$  \\ 
log $\xi$  & 1.57$^{+0.21}_{-0.19}$ \\ 
\hline
\multicolumn{2}{l}{\em $P$}\\
 photon index & 1.91$^{+0.02}_{-0.01}$ \\
$E_{cut}$ (keV) & 160(fixed) \\
\hline
\multicolumn{1}{l}{\em $N_3$}\\
  & $N_1 \times$0.3 \\
\hline
\multicolumn{2}{l}{\em $R^\dagger$}\\
 cosIncl & 0.866(fixed) \\ 
\hline
\multicolumn{2}{l}{\em $I_{Fe}$}\\
line E (keV) & 6.35 (fixed)  \\
sigma (keV) &  0.01 (fixed) \\ 
norm (${10}^{-5}$ph/s/cm$^2$) &  1.2$\pm$0.3 \\ 
EW (eV) &  27$\pm$7  \\ 
\hline
edge E (keV) & 7.11 (fixed) \\
MaxTau & 0.05$\pm$0.02 \\
\hline 
line E (keV) & 7.0 (fixed) \\
sigma (keV) & 0.01(fixed) \\
norm (${10}^{-6}$ph/s/cm$^2$) & $-6.8\pm$2.4 \\
\hline
line E (keV) & 2.35$\pm$0.02 \\
sigma (keV) & 0.01(fixed) \\
norm (${10}^{-5}$ph/s/cm$^2$) & $-2.3\pm$0.5 \\
\hline
reduced chi-square (d.o.f) & 1.12 (203)  \\ 
\hline\hline
\multicolumn{2}{l}{$\dagger$ ``pexrav'' model in xspec is used.}\\
\label{Miyakawa_model_ave_Suzaku}
 \end{tabular}
\end{center}
\end{table}

\setlength{\tabcolsep}{2.4pt}
\begin{table}
\begin{center}
\caption{Results of spectral fitting in 1--40 keV for the average
 spectrum adding Laor's disk line to the three-component model. }
\begin{tabular}{cc}
\hline\hline
\multicolumn{2}{l}{\em Interstellar absorption}\\
$N_H$ (${10}^{21}$ ${\rm cm}^{-2}$)  & 3.5$^{+1.4}_{-0.5}$ \\
\hline
\multicolumn{2}{l}{\em Laor 's disk-line}\\
 E (keV) & 6.40$\pm$0.06  \\
Index & $-3$ (fixed) \\
$r_{in}$ ($R_g$) & 220$^{+180}_{-210}$ \\
$r_{out}$ ($R_g$) & 400 (fixed) \\
inclination (deg) & 30 (fixed) \\
norm (${10}^{-5}$) &  1.3$^{+0.4}_{-0.6}$ \\ 
EW (eV) &  40$\pm$20  \\ 
\hline
\multicolumn{2}{l}{\em $W_H$}\\
$N_H$  (${10}^{23}$ ${\rm cm}^{-2}$) & 2.9$^{+0.9}_{-0.4}$  \\ 
log $\xi$  & 3.35$\pm$0.04 \\ 
\hline
\multicolumn{2}{l}{\em $W_L$}\\
$N_H$  (${10}^{21}$ ${\rm cm}^{-2}$)  & 3.7$^{+1.4}_{-0.8}$ \\ 
log $\xi$  & 1.52$^{+0.21}_{-0.12}$  \\ 
\hline
\multicolumn{2}{l}{\em $N_1$}\\
10$^{-2}$ph/s/cm$^2$ at 1 keV &  1.44$^{+0.04}_{-0.02}$ \\ 
\hline
\multicolumn{2}{l}{\em $N_2$}\\
10$^{-2}$ph/s/cm$^2$ at 1 keV&  0.55$^{+0.05}_{-0.07}$ \\
\hline
\multicolumn{2}{l}{\em $W_2$}\\
$N_H$  (${10}^{24}$ ${\rm cm}^{-2}$)  & 1.7$^{+1.3}_{-0.2}$  \\ 
log $\xi$  & 1.65$^{+0.08}_{-0.16}$ \\ 
\hline
\multicolumn{2}{l}{\em $P$}\\
 photon index & 1.91$^{+0.02}_{-0.01}$ \\
$E_{cut}$ (keV) & 160(fixed) \\
\hline
\multicolumn{2}{l}{\em $N_3$}\\
  & $N_1\times$0.3 \\
\hline
\multicolumn{2}{l}{\em $R^\dagger$}\\
 cosIncl & 0.866(fixed) \\ 
\hline
\multicolumn{2}{l}{\em $I_{Fe}$}\\
line E (keV) & 6.35 (fixed)  \\
sigma (keV) &  0.01 (fixed) \\ 
norm (${10}^{-6}$ph/s/cm$^2$) &  5.4$^{+2.4}_{-4.5}$ \\ 
EW (eV) & 12$^{+5}_{-10}$    \\ 
\hline
edge E (keV) & 7.11 (fixed) \\
MaxTau & 0.04$\pm$0.02 \\
\hline 
line E (keV) & 7.0 (fixed) \\
sigma (keV) & 0.01(fixed) \\
norm (${10}^{-6}$ph/s/cm$^2$) & $-6.1^{+2.1}_{-2.3}$ \\
\hline
line E (keV) & 2.35$\pm$0.02 \\
sigma (keV) & 0.01(fixed) \\
norm (${10}^{-5}$ph/s/cm$^2$) & $-2.3\pm$0.5 \\
\hline
reduced chi-square (d.o.f) &  1.07 (200)  \\ 
\hline\hline
\multicolumn{2}{l}{$\dagger$ ``pexrav'' model in xspec is used.}\\
\label{Miyakawa_diskline_model_ave_Suzaku}
 \end{tabular}
\end{center}
\end{table}

\begin{figure}[htbp]
\begin{center}
\FigureFile(90mm,90mm){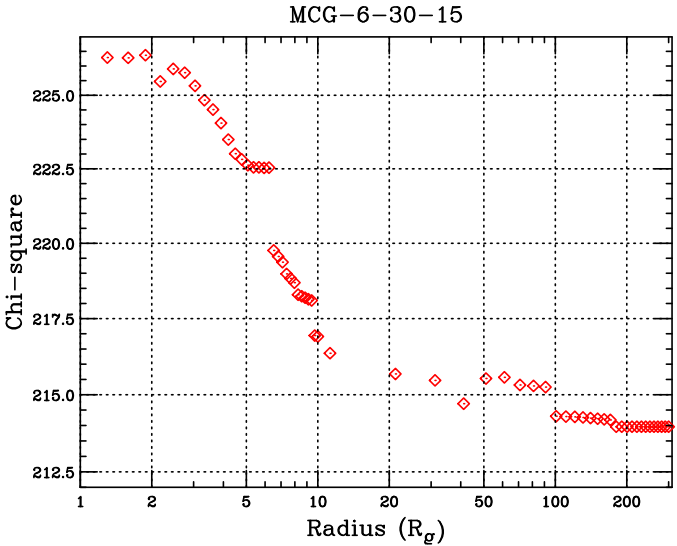}
\end{center}
\caption{Chi-squares of fitting to the average
 spectrum including Laor's disk line model, shown as a function of the inner radius.} 
 \label{Miyakawa_laor_model_fit_Suzaku}
\end{figure}

\subsection{Examination of presence of the ``disk line''}

Several authors claim presence of an extremely broad  ``disk line'' in the
energy spectrum of MCG-6-30-15 (e.g., Miniutti et al.\ 2007).
Therefore,  we examine if presence of the disk line 
is reconciled with our three-component model.
We analyze the time-average spectrum of MCG-6-30-15 adding a disk line
model from a fast-rotating Kerr black hole (Laor 1991). 
Thus, the 
model we adopt here has both the narrow line from very far from the black hole,
and a putative broad line which is  from the region very close to the black hole.
We try to fit the spectrum with Laor`s disk line model with varying  the inner disk
radius, $r_{in}$.

 Table \ref{Miyakawa_diskline_model_ave_Suzaku}
shows the best-fitting result for the Suzaku average spectrum adding Laor's model to
the three-component model, where the fit is slightly improved with 
$r_{in}\approx$ 200 $r_g$ and the 
reduced chi-square is 1.07 (${\chi}^2$/d.o.f = 214/200). 
Note that the additional line is only mildly broadened with
$r_{in}\approx$ 200 $r_g$, and 
the equivalent width of the broad disk line is just 40 $\pm$ 20 eV. 
Figure $\ref{Miyakawa_laor_model_fit_Suzaku}$ shows
the chi-squares of fitting to the average spectrum as a function
of the inner disk radius. As seen from this figure, 
$\Delta$ chi-squares become greater as $r_{in}$ gets less than 
$\approx$ 200 $r_g$, and it exceeds  2.71  
when $r_{in}$ is less than 9 $r_g$.  Namely, our three-component model
is only reconciled with  a mildly 
broad disk line from the region $r_{in} > 9\; r_g$ with 90 \% confidence.

The reason we do not require a strong, extremely distorted disk line 
is primarily because that  the newly introduced heavily
absorbed component in our three-component model (the $N_2 W_2 P$ term in Equation \ref{Eq2})
has an ionized iron edge feature which resembles the broad
disk line shape (Figure \ref{Miyakawa_model_fit_Suzaku}).
Secondary reason is that   the warm absorbers
($W_L$ and $W_H$ in Equation \ref{Eq2}) have spectral curvatures below $\sim$5 keV
without which the residual may look like a low-energy tail of the disk line. 
Consequently, we see  that the disk line feature is dependent on the
choice of the continuum spectral models, and an extremely distorted disk line is not 
required in  our three-component model.


\subsection{Spectral fitting for the sliced spectra}\label{sec:sliced}


Next, we apply the three-component model to the eight ``intensity-sliced spectra''
(the same spectral sets as used in Paper I) to investigate for spectral variations.
The method of creating the intensity-sliced energy  spectra is as follows:
(1) We created a light curve (the average of XIS0, XIS2, and XIS3), with a bin-width of 128 s in the 0.2 -- 12 keV
band.   We found that the counting rate varies  from $\sim 1$ to $\sim 7$
cts s$^{-1}$.  
(2) We chose eight time-periods when the source intensity is in the ranges of  1--1.75, 1.75--2.50,
2.50--3.25, 3.25--4.00, 4.00--4.75, 4.75--5.50, 5.50--6.25, 6.25--7.00 cts s$^{-1}$.
These intensity ranges are chosen so that the exposure time for each intensity bin
be approximately equal. (3) From the eight time-periods corresponding to the different 
source flux levels, we created eight intensity-sliced energy spectra (the sum of XIS0, XIS2,
and XIS3).

Here, all the parameters are made the same for the eight spectra except the 
following three parameters; 
the direct power-law normalization, $N_1$, the absorbed power-law normalization, $N_2$,
and ionization degree of the low-ionized warm absorber, $W_L$.
Attenuation by a  warm absorber is expressed as $W_L \approx \exp (- \sigma(E, \xi) \; N_{H,L})$,
where $\sigma(E, \xi)$ is  energy and ionization-degree dependent cross-section and $N_{H,L}$ 
is the hydrogen column density of the low-ionized absorber.
Variation of $W_L$
is due to either a variation of $\sigma(E, \xi)$ or $N_{H,L}$, or both.
Here, we investigate two cases, one with a variation of the cross-section 
 $\sigma(E, \xi)$ due to change of the ionization parameter $\xi$, the other with a variation of $N_{H, L}$.

Eight spectra are fitted simultaneously, and the 
both cases equally exhibit acceptable fits with reduced chi-square 
values  1.09 and 0.95 (d.o.f = 1135), respectively. 
Thus, we could not tell if  the 
 ionization parameter or column density of the low-ionized warm absorber is more variable from spectral fitting alone.

\begin{figure}[htbp]
\begin{center}
\FigureFile(90mm,90mm){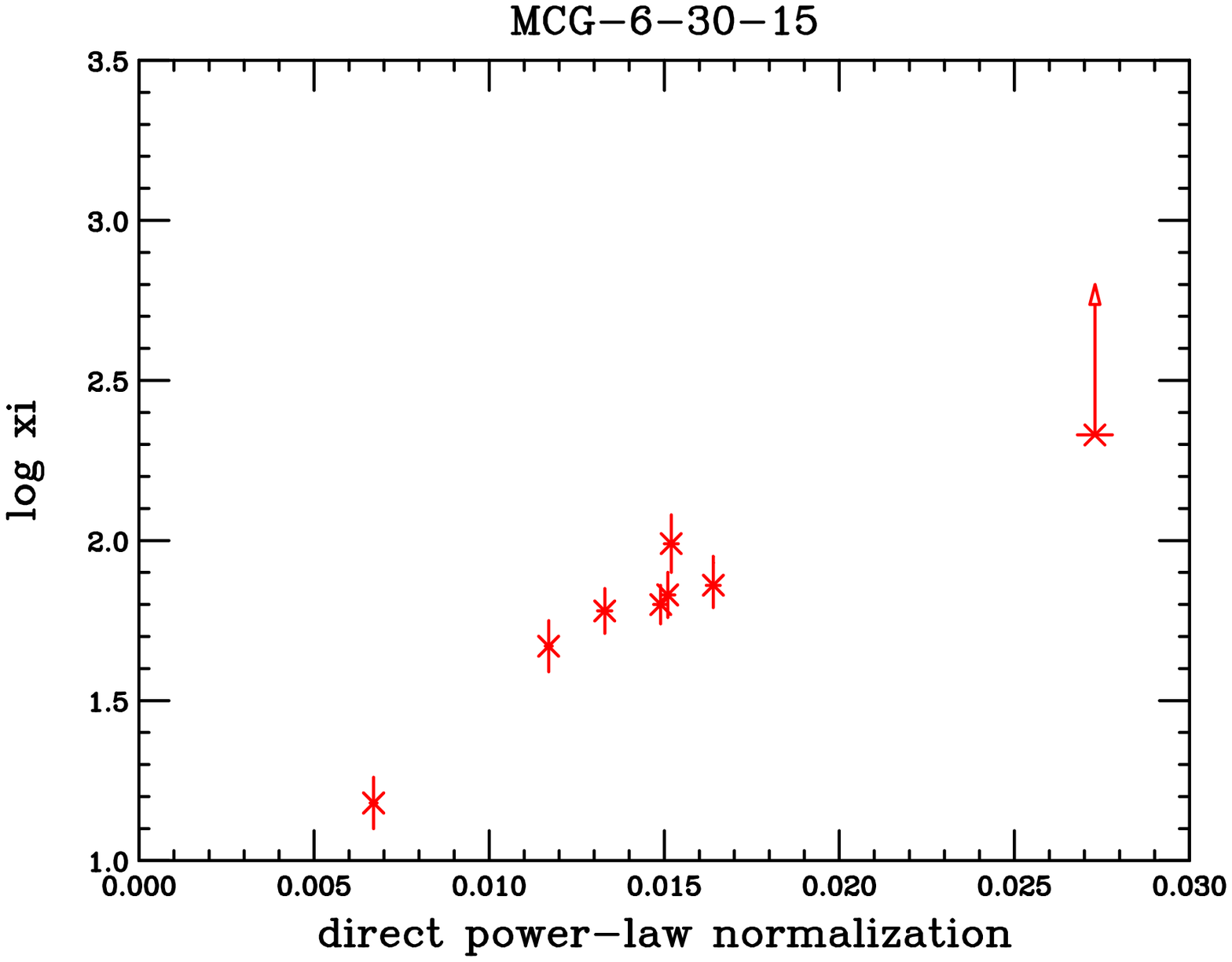}
\FigureFile(90mm,90mm){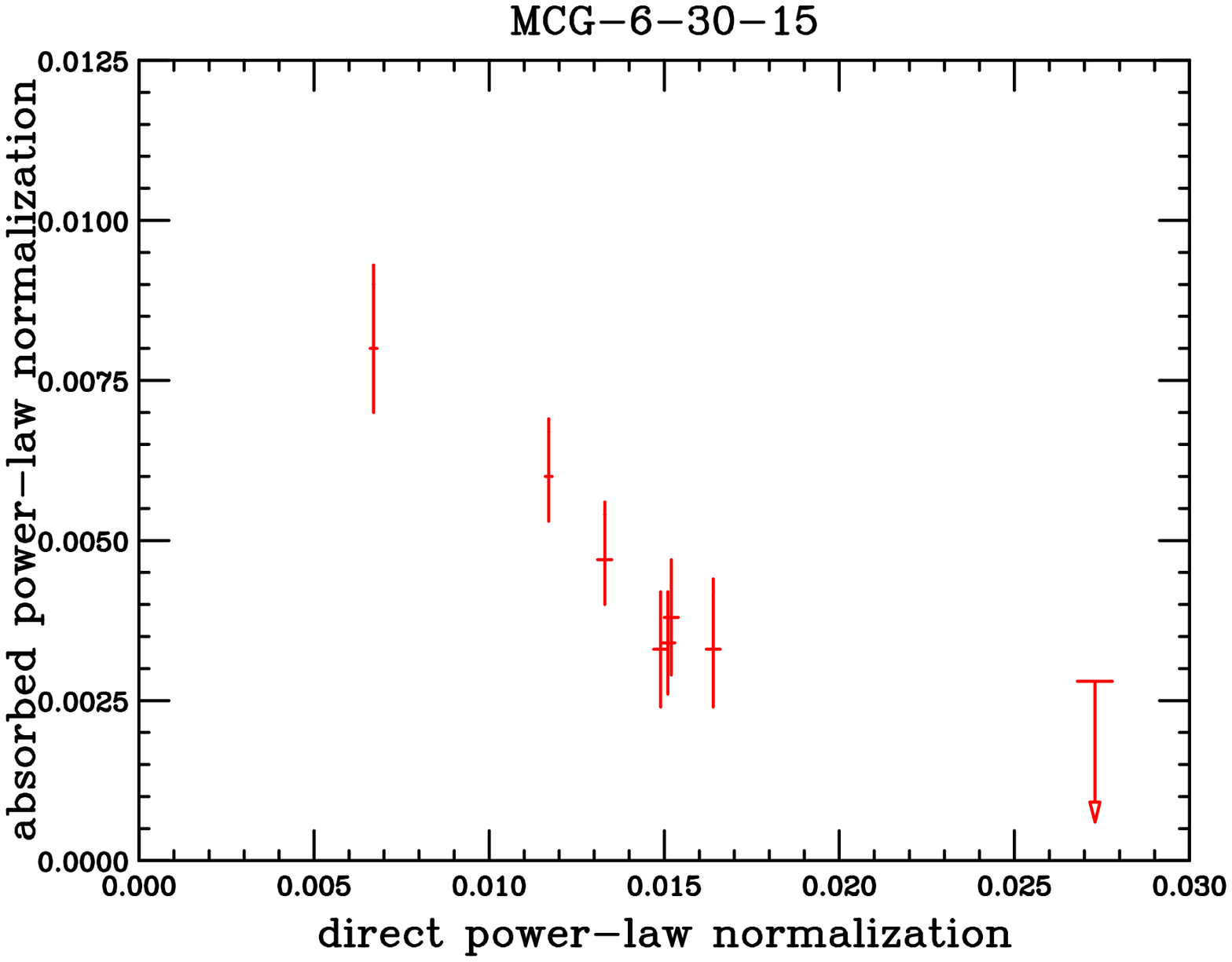}
\end{center}
\caption{Correlation among the three parameters, 
the direct power-law normalization ($N_1$), 
ionization parameter of the low-ionized
warm absorber ($\xi$), and the absorbed power-law normalization ($N_2$),
when applying the three-component model to the eight intensity-sliced spectra.
(Top) Relation between $N_1$ and $\xi$.
(Bottom) Relation between $N_1$ and $N_2$. 
}
 \label{relation_for_the_slice_spectra}
\end{figure}

\begin{figure}[htbp]
\begin{center}
\FigureFile(90mm,90mm){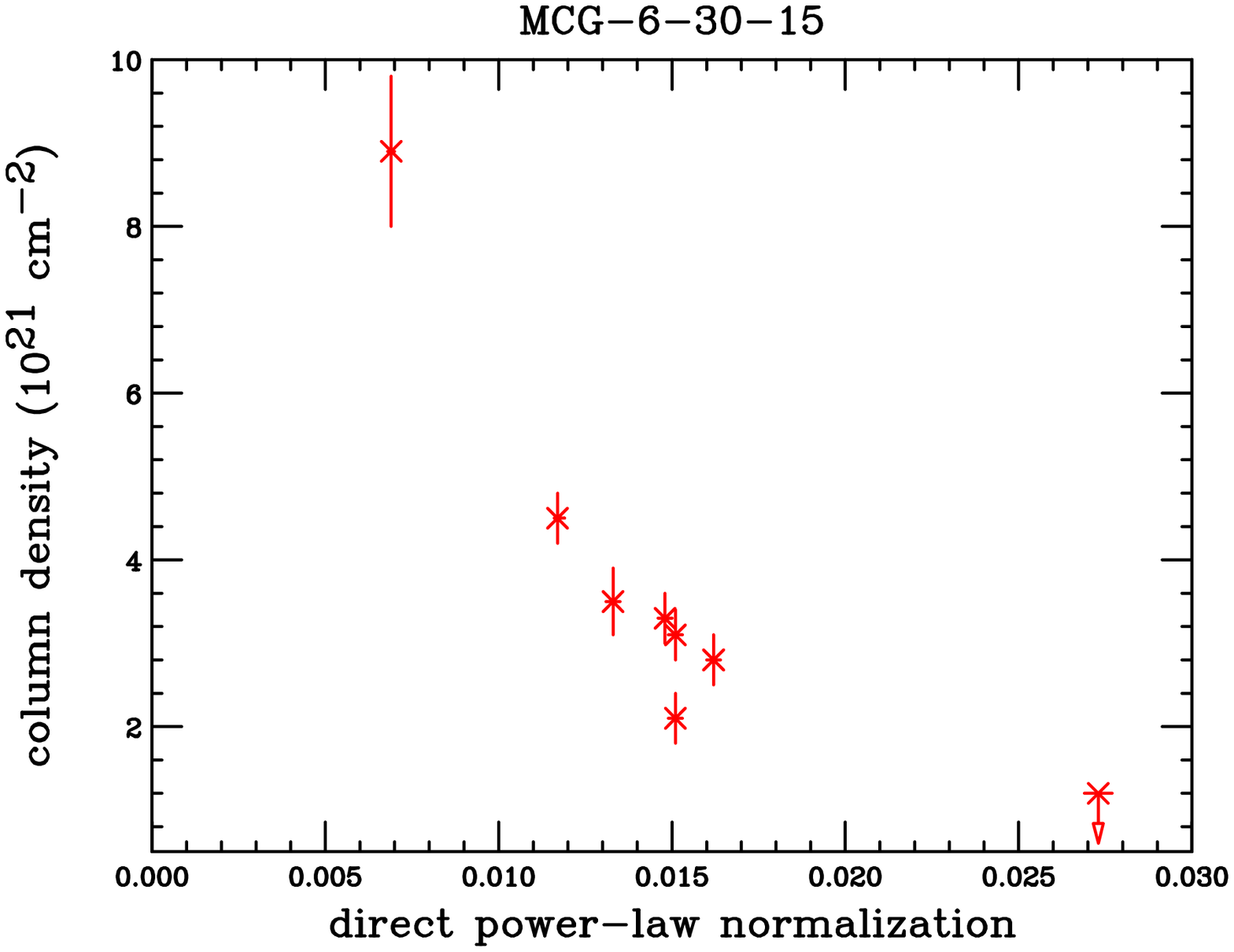}
\FigureFile(90mm,90mm){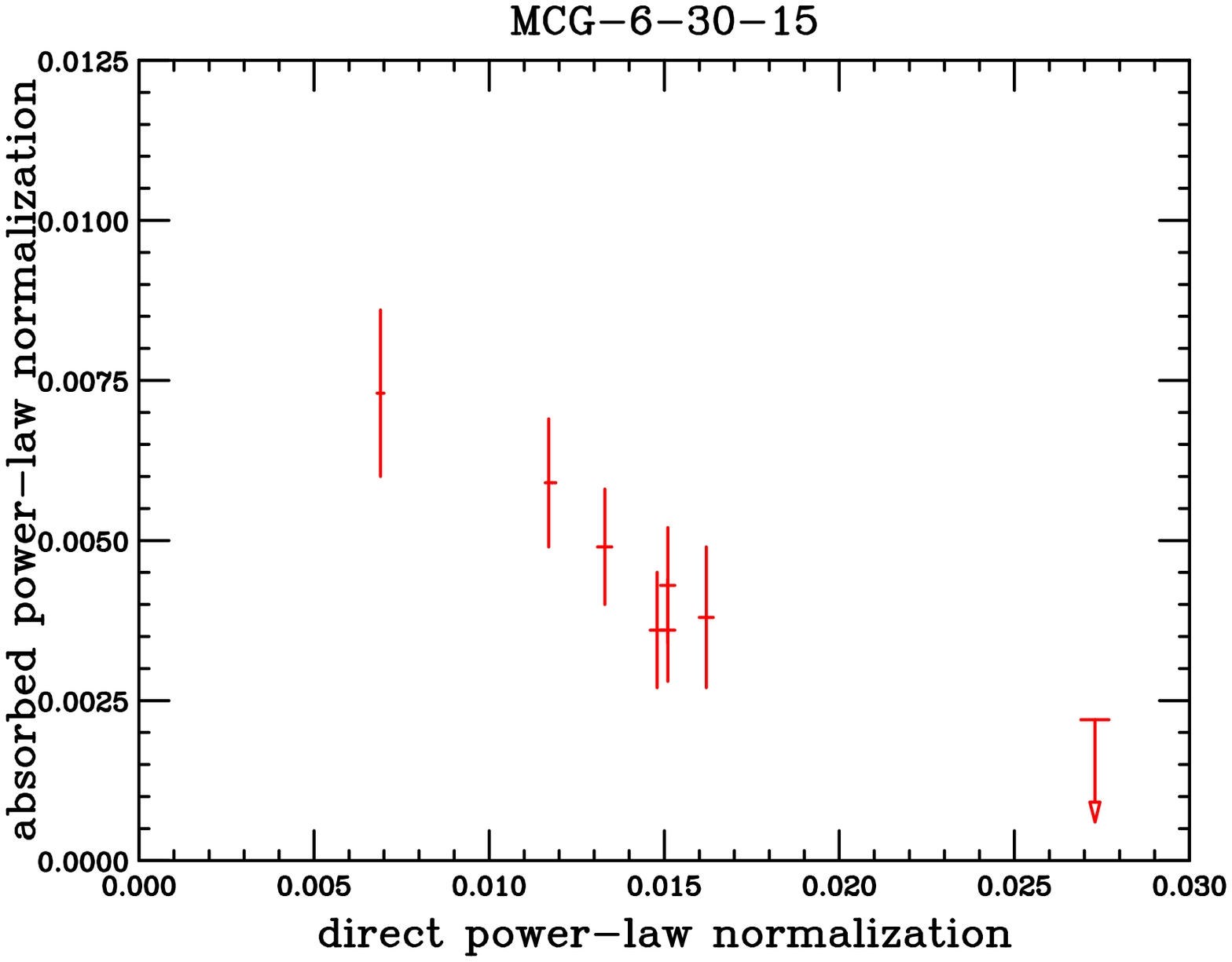}
\end{center}
\caption{Correlation among the three parameters, 
the direct power-law normalization ($N_1$),
column-density  of the low-ionized
warm absorber ($N_{H,L}$), and the absorbed power-law normalization ($N_2$),
when applying the  three-component  model to the eight intensity-sliced spectra.
(Top) Relation between $N_1$ and $N_{H,L}$.
(Bottom) Relation between $N_1$ and $N_2$. }
 \label{relation_for_the_slice_spectra_nH}
\end{figure}

When  $N_{H,L}$ is fixed, the top and bottom panel of Figure $\ref{relation_for_the_slice_spectra}$ shows the relation
 between $N_1$ and $\log \xi$ and  that between $N_1$ and $N_2$, respectively.
Similarly, when   $\xi$ is fixed, the top  and bottom panel of Figure $\ref{relation_for_the_slice_spectra_nH}$ shows the relation
 between $N_1$ and $N_{H,L}$ and that
between $N_1$ and $N_2$, respectively.
In either case,
  correlations between each two of the three
parameters are obvious.  Note that we see  two independent correlations among three free parameters.
This indicates that there should be a single principle parameter which is primarily responsible for
the observed spectral variations (section \ref{sec:vpcmodel}).  

 

\subsection{Spectral fitting for  Chandra/HETGS data}

We have seen that the three-component model is effective  to describe the Suzaku
spectral variation, and that the ionization degree of the low-ionized
warm absorber is variable with intensities.
Next, we will see  if the three-component
model is valid to describe the Chandra HETG spectra and their variations. 
Strength of Chandra HETG compared to Suzaku is that 
it can study variation of individual low-energy absorption lines corresponding to 
change of the ionization degree.

\begin{figure}[htbp]
\begin{center}
\FigureFile(90mm,90mm){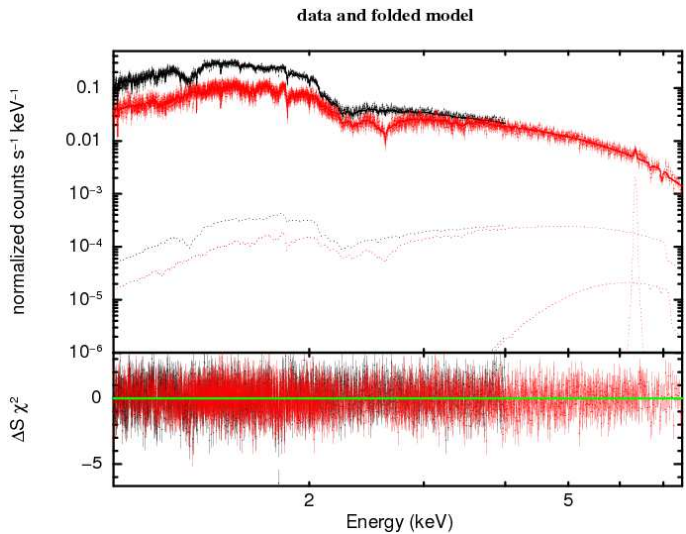}
\end{center}
\caption{Spectral fit result for the time-averaged HETGS spectra with the three-component model. }
 \label{Chandra_average_spec}
\end{figure}

First, 
 we fit the Chandra/HETGS spectra with the three-component model in 1.0 -- 7.5 keV. 
Since the energy range is much smaller than that of Suzaku, some parameters
are not constrained and thus fixed to the Suzaku best-fit values. In  Figure
$\ref{Chandra_average_spec}$ and Table \ref{Miyakawa_model_ave_Chandra},
we show the fitting result for the  average spectrum. We see that the Chandra
HETG spectra are successfully modeled with the three-component model with similar
parameters to those of Suzaku.

Next,  we  study possible ionization degree  variations in the
Chandra/HETGS data by extracting the ``bright spectrum'' and ``faint spectrum''
as follows:
1) Create a light curve with a time-bin-width of 128 sec, and calculate  average counting rates.
2) Create the ``bright spectrum'' from the period when the MEG and HEG 
count rates (0.4--10 keV) are higher than the average, and the ``faint 
spectrum'' when the count rates are lower than the average.

\begin{figure}[htbp]
\begin{center}
\FigureFile(90mm,90mm){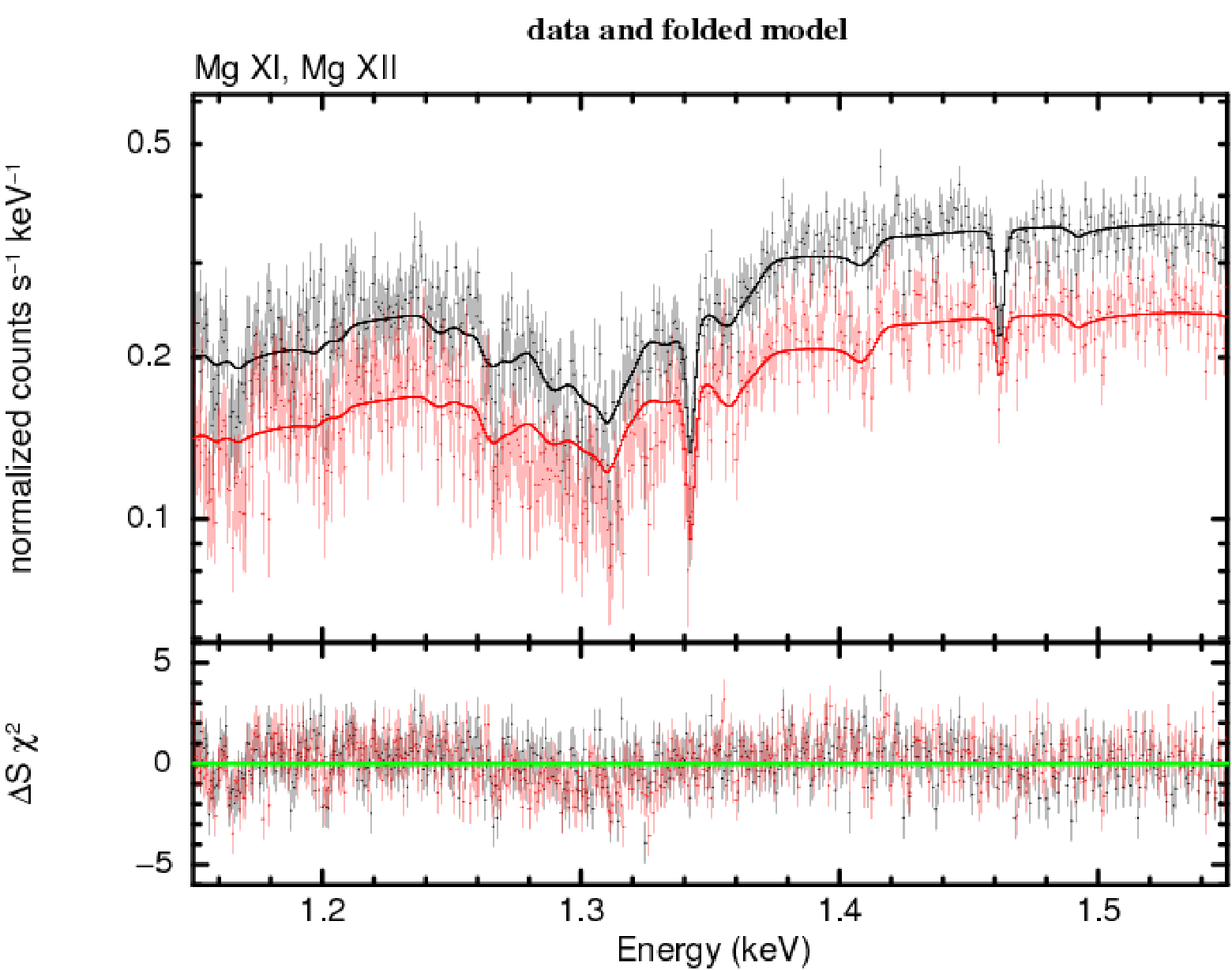}
\FigureFile(90mm,90mm){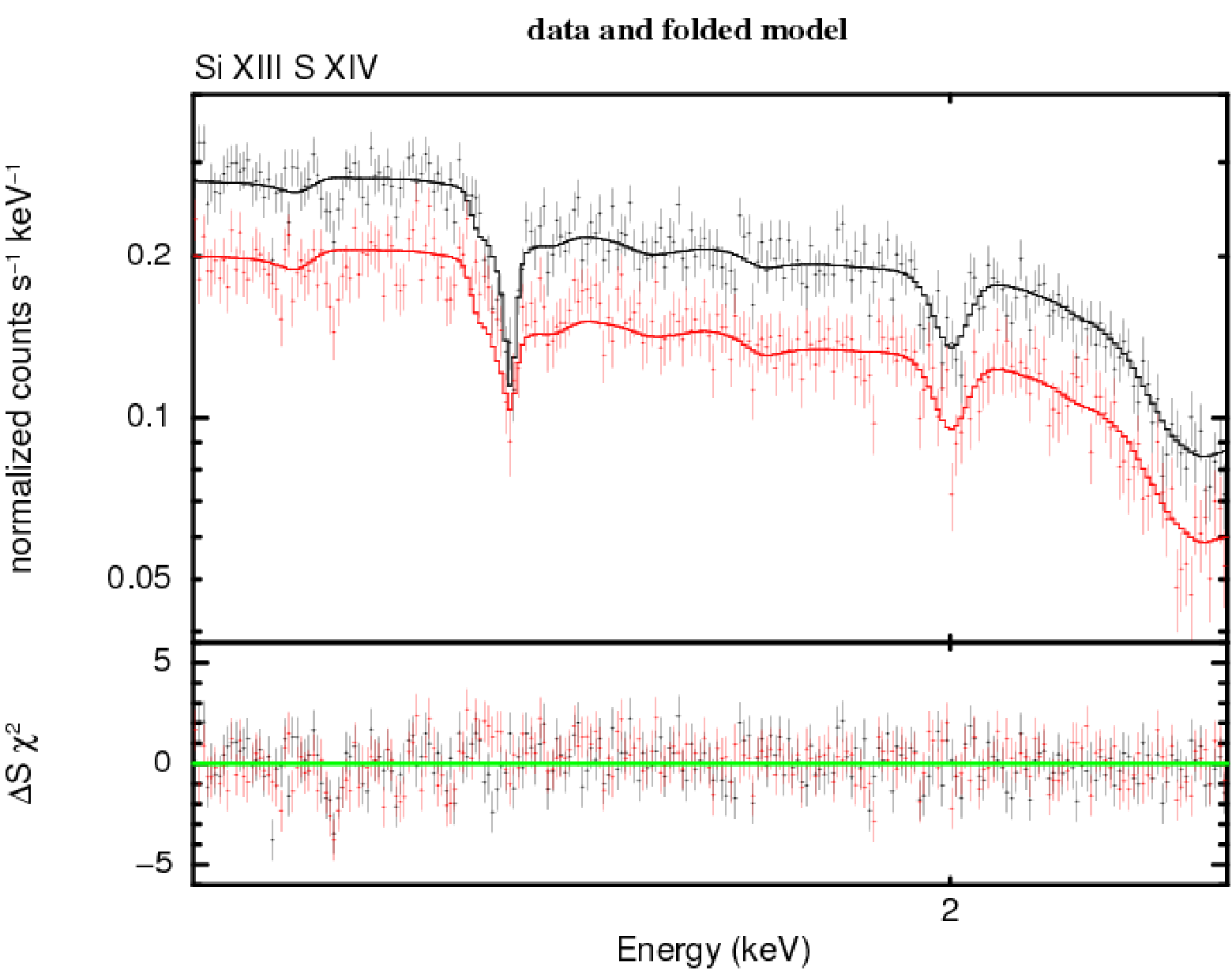}
\end{center}
\caption{(Top) Chandra HETGS simultaneous fit in the 1.15--1.55 keV band for the bright
 spectrum and faint spectrum with a power-law plus two negative gaussian model.
 Mg absorption lines are at 1.34 keV and 1.47 keV.
(Bottom) Simultaneous fit in the 1.75--2.10 keV band with a power-law plus two negative gaussian model. Si absorption lines are at 1.85 keV and 2.00 keV.}
\label{around_Si} \label{around_Mg}
\end{figure}

We fit the bright/faint spectra in the 1.15--1.55 keV band
with a single power-law and two negative gaussians for
 the  Mg{\footnotesize XI} (He-like 1.34 keV) and  
Mg{\footnotesize XII} (H-like 1.47 keV) absorption lines.
We also fit the bright/faint spectra in the 1.75--2.10 keV band separately 
with a single power-law and two negative gaussians to
investigate for the Si{\footnotesize XIII} (He-like 1.85--1.86 keV) and  
Si{\footnotesize XIV} (H-like 2.00 keV) lines.
In Figure $\ref{around_Mg}$, we show the fitting result for the bright/faint 
spectra in both energy bands.
We could  fit the bright
and faint spectra with the common photon index, while only the power-law
normalization and the normalization of absorption lines are varied.
The best fit parameters are shown in Table $\ref{table_chandra}$.

We notice that the 
Mg{\footnotesize XI} absorption line equivalent width is
larger in the faint state, while that of the 
Mg{\footnotesize XII} line is larger in the bright state.
This is understood as  due to change of the ionization state
of the warm absorber, such that 
Mg{\footnotesize XI} is more abundant in the faint state,
while Mg{\footnotesize XII} is more abundant in the bright state.
In fact, ion fraction of Mg{\footnotesize XI} and 
that of Mg{\footnotesize XII} become equal at $\log \xi 
\simeq 2$ (e.g., Kallman $\&$ Bautista 2001), that is about  the best-fit
value of the low-ionized warm absorber (Tables \ref{tab:chandra3comp}).

As for Si absorption lines,
the Si{\footnotesize XIII} 
line equivalent width is oppositely larger for the bright spectrum,
which is explained by the low-ionized  warm absorber with 
$\log \xi \simeq 2$, since ion fraction of Si{\footnotesize XIII} 
increases  when $\log \xi$ varies from 
$\sim1.5$ to $\sim2$ (Kallman $\&$ Bautista 2001).
Fraction of Si{\footnotesize XIV} 
should increase more dramatically when $\log \xi$ increases accordingly
 (Kallman $\&$ Bautista 2001), but the observed equivalent widths
  of the Si{\footnotesize XIV}  line is not much different.
This result is explained by the presence of the high-ionized 
warm absorber at $\log \xi \geq 3$, which produces a significant amount
of  Si{\footnotesize XIV} but not  Si{\footnotesize XIII}. 
Consequently, observed variation of the Mg and Si absorption line equivalent-widths
 is primarily explained by change of the ionization 
degree of the low-ionized warm absorber 
according to the flux changes.


\setlength{\tabcolsep}{2.4pt}
\begin{table}
\begin{center}\label{tab:chandra3comp}
\caption{Results of spectral fitting of Chandra/HETG in 1--7.5 keV with the three-component model
for the average spectrum. }
\begin{tabular}{cc}
\hline\hline
\multicolumn{2}{l}{\em Interstellar absorption}\\
$N_H$ (${10}^{21}$ ${\rm cm}^{-2}$)  & 1.5(fixed) \\
\hline
\multicolumn{2}{l}{\em $W_H$}\\
$N_H$  (${10}^{23}$ ${\rm cm}^{-2}$) & 2.4(fixed)  \\ 
log $\xi$  & 3.64$\pm$0.04 \\ 
redshift (${10}^{-3}$) &  1.8$^{+0.2}_{-0.5}$ \\
\hline
\multicolumn{2}{l}{\em $W_L$}\\
$N_H$  (${10}^{21}$ ${\rm cm}^{-2}$)  & 3.7(fixed) \\ 
log $\xi$  & 2.04$\pm$0.03  \\
redshift (${10}^{-3}$) &  7.4$\pm$0.1\\
\hline
\multicolumn{2}{l}{\em $N_1$}\\
10$^{-2}$ ph/s/cm$^2$ at 1 keV&  1.23$\pm$0.01 \\ 
\hline
\multicolumn{2}{l}{\em $N_2$}\\
10$^{-2}$ ph/s/cm$^2$ at 1 keV&  $<$0.2 \\
\hline
\multicolumn{2}{l}{\em $W_2$}\\
$N_H$  (${10}^{24}$ ${\rm cm}^{-2}$)  & 1.6(fixed)  \\ 
log $\xi$  & 1.57(fixed) \\
redshift (${10}^{-3}$) & 1(fixed) \\
\hline
\multicolumn{2}{l}{\em $P$}\\
 photon index & 1.76$\pm$0.01 \\
$E_{cut}$ (keV) & 160(fixed) \\
\hline
\multicolumn{1}{l}{\em $N_3$}\\
  & $N_1 \times$0.3 \\
\hline
\multicolumn{2}{l}{\em $R^\dagger$}\\
 cosIncl & 0.866(fixed) \\ 
\hline
\multicolumn{2}{l}{\em $I_{Fe}$}\\
line E (keV) & 6.34 (fixed)  \\
sigma (keV) &  0.018 (fixed) \\ 
norm (${10}^{-5}$ph/s/cm$^2$) &  1.2$\pm$0.4 \\ 
EW (eV) &  25$\pm$8  \\ 
\hline
edge E (keV) & 7.11 (fixed) \\
MaxTau & 0.13$^{+0.10}_{-0.09}$ \\
\hline
line E (keV) & 6.98$^{+0.03}_{-0.06}$  \\
sigma (keV) & 0.01(fixed) \\
norm (${10}^{-6}$ ph/s/cm$^2$) & $-5.0^{+3.9}_{-7.6}$ \\
\hline
reduced chi-square (d.o.f) &  0.92 (5099)  \\ 
\hline\hline
\multicolumn{2}{l}{$\dagger$ ``pexrav'' model in xspec is used.}\\
\label{Miyakawa_model_ave_Chandra}
\end{tabular}
\end{center}
\end{table}

\begin{table}
\begin{center}
\caption{Results of spectral fitting of Chandra/HETGS spectra in 1.15--1.55 keV  
and 1.75 -- 2.10 keV for the intensity
 sliced spectra when only the power-law normalization and the
 normalization of absorption lines are varied.}
\begin{tabular}{lccc}
\hline
\hline
 & Bright & Faint  \\
\hline
\multicolumn{3}{c}{Mg XI absorption line} \\
line E (keV) & \multicolumn{2}{c}{1.3424$\pm$0.0004} \\
sigma (10$^{-3}$ keV) &  \multicolumn{2}{c}{1.4$\pm$0.5}\\
norm  (10$^{-5}$ph/s/cm$^2$ ) &  $-1.2\pm$0.3 &  $-1.0\pm$0.2  \\
 EW (eV) & $-1.9\pm$0.5 & $-2.2\pm$0.4 \\ 
\hline
\multicolumn{3}{c}{Mg XII absorption line} \\
line E (keV) & \multicolumn{2}{c}{1.4621$\pm$0.0004} \\
sigma (10$^{-3}$ keV) &  \multicolumn{2}{c}{1.2$\pm$0.6} \\
norm  (10$^{-5}$ph/s/cm$^2$) &  $-1.0\pm$0.2 &  $-0.4\pm$0.2  \\
EW (eV) & $-1.7\pm$0.3  & $-1.0\pm$0.5  \\
\hline
photon index & \multicolumn{2}{c}{0.95$\pm$0.07} \\
norm (10$^{-3}$ph/s/cm$^2$ at 1 keV ) & 8.6$\pm$0.2 & 6.0$\pm$0.2 \\
\hline
reduced chi-square & \multicolumn{2}{c}{1.21 (${\chi}^2$/d.o.f = 1329/1096)} \\
\hline 
\hline
\hline
 & Bright & Faint  \\
\hline
\multicolumn{3}{c}{Si XIII absorption line} \\
line E (keV) & \multicolumn{2}{c}{1.8515$\pm$0.0005} \\
sigma ( 10$^{-3} $keV) &  \multicolumn{2}{c}{ $<$ 1.7}\\
norm  ( 10$^{-5}$ ph/s/cm$^2$) &  $-1.0\pm$0.2 &  0.4$\pm$0.2  \\
 EW (eV) & $-2.4\pm$0.5 &  $-1.4\pm$0.7  \\ 
\hline
\multicolumn{3}{c}{Si XIV absorption line} \\
line E (keV) & \multicolumn{2}{c}{2.000$\pm$0.002} \\
sigma ( 10$^{-3} $keV) &  \multicolumn{2}{c}{5.5$^{+1.9}_{-2.1}$} \\
norm  ( 10$^{-5}$ ph/s/cm$^2$) &  $-1.5\pm$0.4  &  $-1.1\pm$0.4   \\
EW (eV) &  $-3.8\pm$1.0  &  $-3.9\pm$1.4  \\
\hline
photon index & \multicolumn{2}{c}{0.97$\pm$0.15} \\
norm ( 10$^{-3}$ ph/s/cm$^2$ at 1 keV ) &  7.7$\pm$0.8 & 5.7$\pm$0.6 \\
\hline
reduced chi-square & \multicolumn{2}{c}{1.17 (${\chi}^2$/d.o.f = 539/459)} \\
\hline 
\label{table_chandra}\label{table_chandra2}
\end{tabular}
\end{center}
\end{table}

\setlength{\tabcolsep}{2.4pt}
\begin{table}
\begin{center}
\caption{Results of spectral fitting of Chandra/HETG in 1--7.5 keV with the three-component model
for the Bright and Faint  spectra varying the {\em ionization parameter}\/ of the low-ionized
warm absorber. }
\begin{tabular}{ccc}
\hline\hline
\hline
&Bright & Faint \\
\hline
\multicolumn{2}{l}{\em Interstellar absorption}\\
$N_H$ (${10}^{21}$ ${\rm cm}^{-2}$)  & \multicolumn{2}{c}{1.5(fixed)} \\
\hline
\multicolumn{2}{l}{\em $W_H$}\\
$N_H$  (${10}^{23}$ ${\rm cm}^{-2}$) &\multicolumn{2}{c}{2.4(fixed)}  \\ 
log $\xi$  & 3.64$\pm$0.05 & $3.62\pm0.06$ \\ 
redshift (${10}^{-3}$) &  \multicolumn{2}{c}{1.82$^{+0.24}_{-0.29}$} \\
\hline
\multicolumn{2}{l}{\em $W_L$}\\
$N_H$  (${10}^{21}$ ${\rm cm}^{-2}$)  & \multicolumn{2}{c}{3.7(fixed)} \\ 
log $\xi$  & 1.95$\pm$0.04 & 1.78$\pm$0.04  \\
redshift (${10}^{-3}$) &  \multicolumn{2}{c}{7.4$\pm^{0.2}_{0.1}$}\\
\hline
\multicolumn{2}{l}{\em $N_1$}\\
10$^{-2}$ ph/s/cm$^2$ at 1 keV&  1.47$\pm$0.01&$1.05\pm0.01$ \\ 
\hline
\multicolumn{2}{l}{\em $N_2$}\\
10$^{-2}$ ph/s/cm$^2$ at 1 keV&  \multicolumn{2}{c}{$1.5-N_1$}  \\
\hline
\multicolumn{2}{l}{\em $W_2$}\\
$N_H$  (${10}^{24}$ ${\rm cm}^{-2}$)  &\multicolumn{2}{c}{1.6(fixed)}  \\ 
log $\xi$  & \multicolumn{2}{c}{1.57(fixed)} \\
redshift (${10}^{-3}$) & \multicolumn{2}{c}{1(fixed)} \\
\hline
\multicolumn{2}{l}{\em $P$}\\
 photon index & \multicolumn{2}{c}{1.78$\pm$0.04} \\
$E_{cut}$ (keV) & \multicolumn{2}{c}{160(fixed)} \\
\hline
\multicolumn{1}{l}{\em $N_3$}\\
  & \multicolumn{2}{c}{$N_1 \times$0.3} \\
\hline
\multicolumn{2}{l}{\em $R^\dagger$}\\
 cosIncl & \multicolumn{2}{c}{0.866(fixed)} \\ 
\hline
\multicolumn{2}{l}{\em $I_{Fe}$}\\
line E (keV) & \multicolumn{2}{c}{6.34 (fixed)}  \\
sigma (keV) &  \multicolumn{2}{c}{0.018 (fixed)} \\ 
norm (${10}^{-6}$ph/s/cm$^2$) &  \multicolumn{2}{c}{7.8$\pm$4.3} \\ 
EW (eV) &  14$\pm$8 & $18\pm10$  \\ 
\hline
edge E (keV) & \multicolumn{2}{c}{7.11 (fixed)} \\
MaxTau & \multicolumn{2}{c}{0.13$^{+0.10}_{-0.09}$} \\
\hline
line E (keV) & \multicolumn{2}{c}{6.98$^{+0.03}_{-0.06}$}  \\
sigma (keV) & \multicolumn{2}{c}{0.01(fixed)} \\
norm (${10}^{-6}$ ph/s/cm$^2$) & \multicolumn{2}{c}{$-5.0^{+3.9}_{-7.6}$} \\
\hline
reduced chi-square (d.o.f) &  \multicolumn{2}{c}{0.89 (7505)}  \\ 
\hline\hline
\multicolumn{2}{l}{$\dagger$ ``pexrav'' model in xspec is used.}\\
\label{Miyakawa_model_slice_Chandra}
\end{tabular}
\end{center}
\end{table}

\setlength{\tabcolsep}{2.4pt}
\begin{table}
\begin{center}
\caption{Results of spectral fitting of Chandra/HETG in 1--7.5 keV with the three-component model
for the Bright and Faint  spectra varying the {\em column-density}\/  of the low-ionized
warm absorber. }
\begin{tabular}{ccc}
\hline\hline
\hline
&Bright & Faint \\
\hline
\multicolumn{2}{l}{\em Interstellar absorption}\\
$N_H$ (${10}^{21}$ ${\rm cm}^{-2}$)  & \multicolumn{2}{c}{1.5(fixed)} \\
\hline
\multicolumn{2}{l}{\em $W_H$}\\
$N_H$  (${10}^{23}$ ${\rm cm}^{-2}$) & \multicolumn{2}{c}{2.4 (fixed)}\\
log $\xi$  & 3.65$\pm$0.05 & 3.62$\pm$0.06 \\
redshift (${10}^{-3}$) &  \multicolumn{2}{c}{1.82$^{+0.23}_{-0.31}$} \\
\hline
\multicolumn{2}{l}{\em $W_L$}\\
$N_H$  (${10}^{21}$ ${\rm cm}^{-2}$) &3.4$\pm$0.4 & 3.5$\pm$0.4  \\ 
log $\xi$  & \multicolumn{2}{c}{1.54 (fixed)}\\
redshift (${10}^{-3}$) &  \multicolumn{2}{c}{7.4$\pm^{0.2}_{0.1}$}\\
\hline
\multicolumn{2}{l}{\em $N_1$}\\
10$^{-2}$ ph/s/cm$^2$ at 1 keV&  1.57$\pm$0.03&$1.11\pm0.02$ \\ 
\hline
\multicolumn{2}{l}{\em $N_2$}\\
10$^{-2}$ ph/s/cm$^2$ at 1 keV&  \multicolumn{2}{c}{$1.6-N_1$}  \\
\hline
\multicolumn{2}{l}{\em $W_2$}\\
$N_H$  (${10}^{24}$ ${\rm cm}^{-2}$)  &\multicolumn{2}{c}{1.6(fixed)}  \\ 
log $\xi$  & \multicolumn{2}{c}{1.57(fixed)} \\
redshift (${10}^{-3}$) & \multicolumn{2}{c}{1(fixed)} \\
\hline
\multicolumn{2}{l}{\em $P$}\\
 photon index & \multicolumn{2}{c}{1.83$\pm^{0.02}_{0.01}$} \\
$E_{cut}$ (keV) & \multicolumn{2}{c}{160(fixed)} \\
\hline
\multicolumn{1}{l}{\em $N_3$}\\
  & \multicolumn{2}{c}{$N_1 \times$0.3} \\
\hline
\multicolumn{2}{l}{\em $R^\dagger$}\\
 cosIncl & \multicolumn{2}{c}{0.866(fixed)} \\ 
\hline
\multicolumn{2}{l}{\em $I_{Fe}$}\\
line E (keV) & \multicolumn{2}{c}{6.34 (fixed)}  \\
sigma (keV) &  \multicolumn{2}{c}{0.018 (fixed)} \\ 
norm (${10}^{-6}$ph/s/cm$^2$) &  \multicolumn{2}{c}{9.1$\pm$4.3} \\ 
EW (eV) &  17$\pm$8 & $22\pm10$  \\ 
\hline
edge E (keV) & \multicolumn{2}{c}{7.11 (fixed)} \\
MaxTau & \multicolumn{2}{c}{0.13$^{+0.11}_{-0.09}$} \\
\hline
line E (keV) & \multicolumn{2}{c}{6.98$^{+0.04}_{-0.07}$}  \\
sigma (keV) & \multicolumn{2}{c}{0.01(fixed)} \\
norm (${10}^{-6}$ ph/s/cm$^2$) & \multicolumn{2}{c}{$-10.3^{+7.3}_{-7.7}$} \\
\hline
reduced chi-square (d.o.f) &  \multicolumn{2}{c}{0.91 (7505)}  \\ 
\hline\hline
\multicolumn{2}{l}{$\dagger$ ``pexrav'' model in xspec is used.}\\
\label{Miyakawa_model_slice_Chandra_columnb}
\end{tabular}
\end{center}
\end{table}


 We further simultaneously fit the bright/faint spectra in the 1.0--7.5 keV band with the three-component model
 (Table $\ref{Miyakawa_model_slice_Chandra}$ and \ref{Miyakawa_model_slice_Chandra_columnb})
by varying normalization of the direct component and ionization degree of the low-ionized 
warm absorber.
 In stead of allowing the normalization of the absorbed power-law
 component to be free, we fixed it to the value of a constant total  
 normalization minus the direct power-law normalization, in order to 
 confirm the anti-correlation between
 the direct component and   absorbed component  as observed
in Suzaku (Bottom panels 
 in Figures \ref{relation_for_the_slice_spectra} and \ref{relation_for_the_slice_spectra_nH}).
In Table $\ref{Miyakawa_model_slice_Chandra}$, we show results of allowing  the ionization
degrees of the low-ionized warm absorber to be free while the column density is fixed;
we see the parameter is  significantly variable
 ($\log \xi$ = 1.95$\pm$0.04 and 1.78$\pm$0.04).
Also, ionization degree of the high-ionized warm absorber is made a free parameter,
but  these values are hardly variable   ($\log \xi$ = 3.64$\pm$0.05 and 3.62$\pm$0.06).

Because of the presence of constant high-ionized warm absorber, varying  column density
of the low-ionized  warm absorber effectively changes the average ionization states.
Therefore, in stead of varying  ionization degree of the low-ionized warm absorber,
the bright and faint spectral variation  can be equally explained by 
varying its column density.  This result is shown in
Table \ref{Miyakawa_model_slice_Chandra_columnb}.

In summary, Chandra/HETGS data confirm that 
 the spectral  variation 
 is described by  change of only two spectral parameters
in the framework of the three-component model (Equation \ref{Eq2}), 
 normalization of the direct component ($N_1$) and either ionization degree or column-density
of the low-ionized warm absorber($W_L$). 







\subsection{Time-variation of the spectral parameters  and introduction of  the ``covering fraction''}

Next, we create a time-sequence of the energy spectra from Suzaku data, 
and apply the three-component model.
We have  made energy spectra every 20 ksec, and performed spectral fits to all the sequential 20 ksec spectra with the three-component model, 
to investigate time variation of the spectral parameters. 
We chose a 20 ksec interval, 
 since variation amplitudes in the 1--10 keV band is maximum at around 20--80 ksec
 (Inoue, Miyakawa and Ebisawa  2011).

In the spectral fits, 
$N_1$, $N_2$ and $W_L$ are treated as time-variable in Equation (\ref{Eq2}).
The fits are made
for the three observation sequences separately
(obsID:700007010, 700007020 and  700007030).
 We have seen that observed variation of $W_L$ is equally represented with change of
 the column density $N_{H, L}$ or  ionization parameter, $\xi$ (section \ref{sec:sliced}).
 Hereafter, we vary  $N_{H, L}$ while $\xi$ is assumed to be invariable throughout the observation.  In fact, as we shall see below, variation 
 of $N_{H, L}$ is  more physically reasonable than that of $\xi$.

\begin{figure}[htbp]
\FigureFile(90mm,90mm){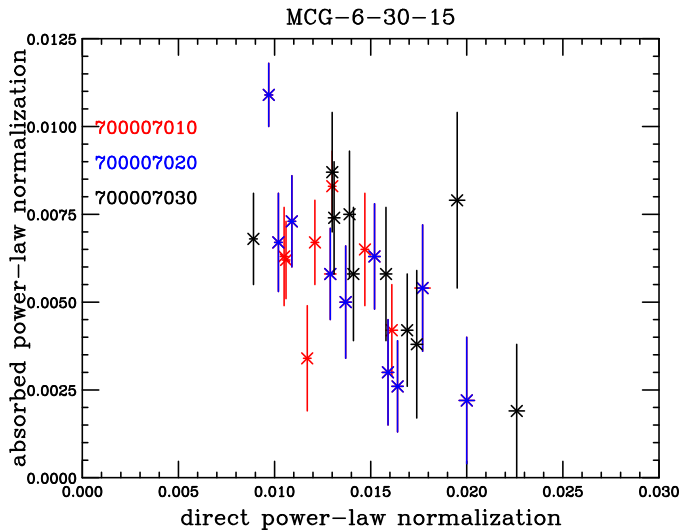}
\caption{Correlation between normalizations
of the direct component and the absorbed component with the
fitting with every 20 ksec.}
 \label{direct_absorbed_20ksec}
\end{figure}

Figure \ref{direct_absorbed_20ksec} shows a correlation between $N_1$ (direct power-law normalization) and $N_2$  (absorbed power-law normalization), 
where an inverse correlation is  seen between them, as well as in the
bottom panels in
 Figure \ref{relation_for_the_slice_spectra} and \ref{relation_for_the_slice_spectra_nH}.
The correlation coefficient between $N_1$ and $N_2$ is $-0.66$. If we fit with a linear relation, $N_2 = a \times N_1 + b$,
we obtain $a=-0.58$, $b=0.01$  and $\chi^2_r = 1.53$ (d.o.f.=35).  On the other hand, fits with a constant $N_1$ 
($N_1=0.013$ with $\chi^2_r = 139.2 $ for d.o.f.=36) or $N_2$ ($N_2=0.0063, \chi^2_r = 3.18$ for d.o.f.=36) 
 are not acceptable.

   We found  that  slope of the anti-correlation between $N_1$ and $N_2$ is not very far from $-1$, which suggests 
that sum of the two parameters may be rather constant.
 Therefore, we introduce the 
  ``total normalization'', $N$, as,  
\begin{equation}
 N \equiv N_1 + N_2.
\label{Eq3}
\end{equation}
 Similarly, we introduce the ``covering fraction'', $\alpha$, as 
\begin{equation}
 \alpha \equiv \frac{ N_2 } { N }.
\label{Eq4}
\end{equation}

\begin{figure}[htbp]
\FigureFile(90mm,90mm){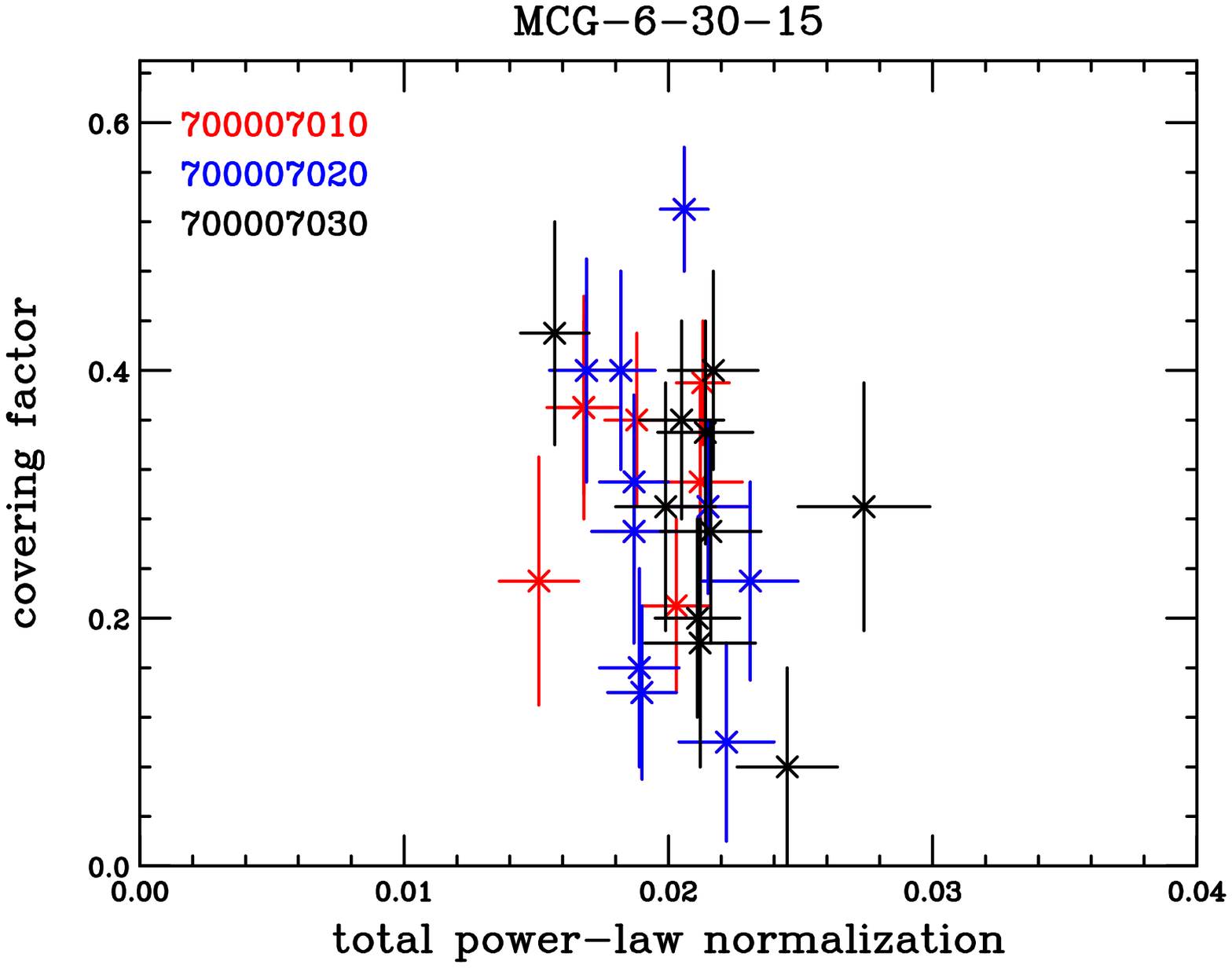}
\FigureFile(90mm,90mm){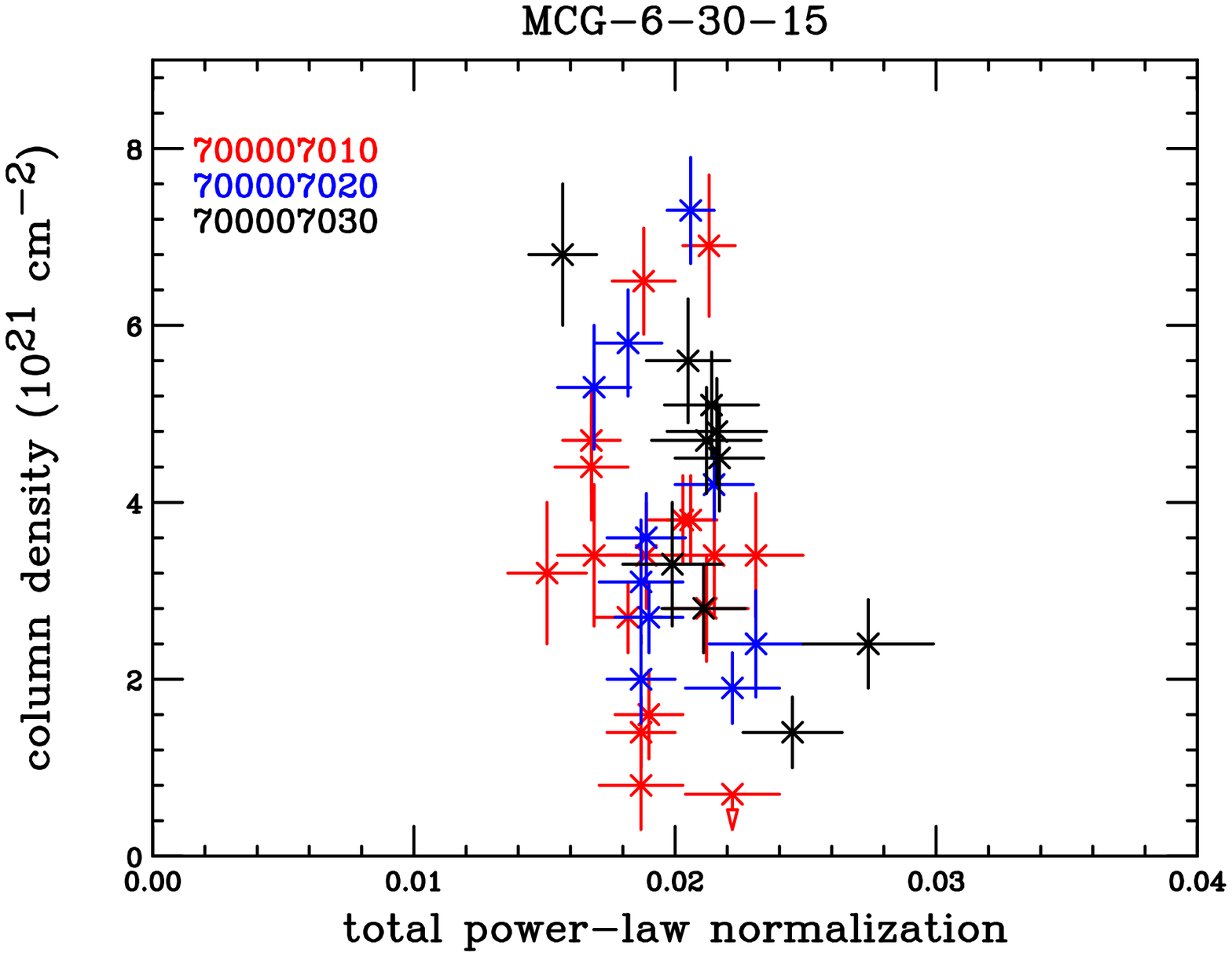}
\FigureFile(90mm,90mm){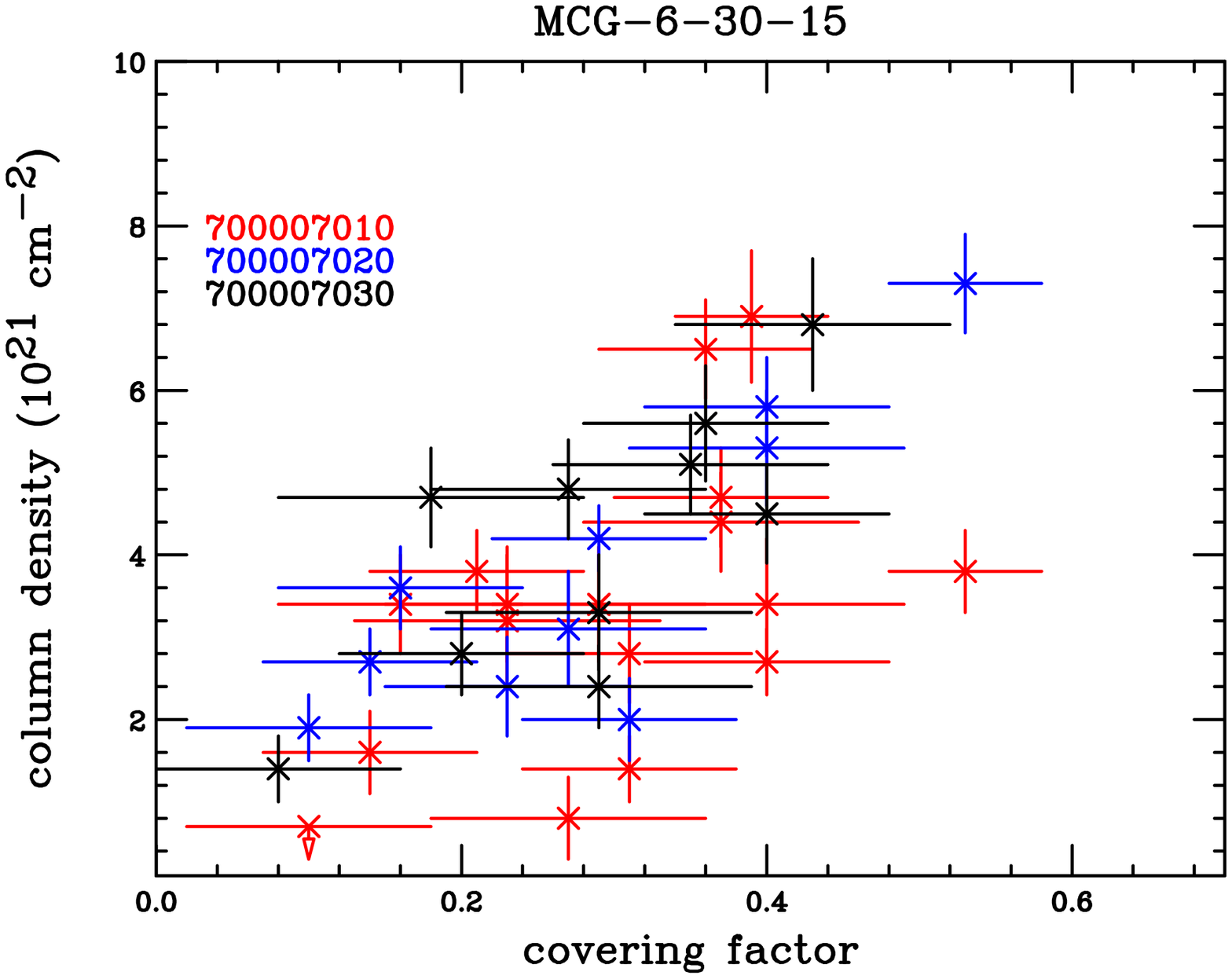}
\caption{Correlation among the covering fraction, $\alpha$, total normalization, $N$,  
and column density of the low-ionized warm absorber, $N_{H,L}$ for the spectral fitting for
every 20 ksec.
}
 \label{correlation_20ksec}
\end{figure}

\begin{figure}[htbp]
\FigureFile(90mm,90mm){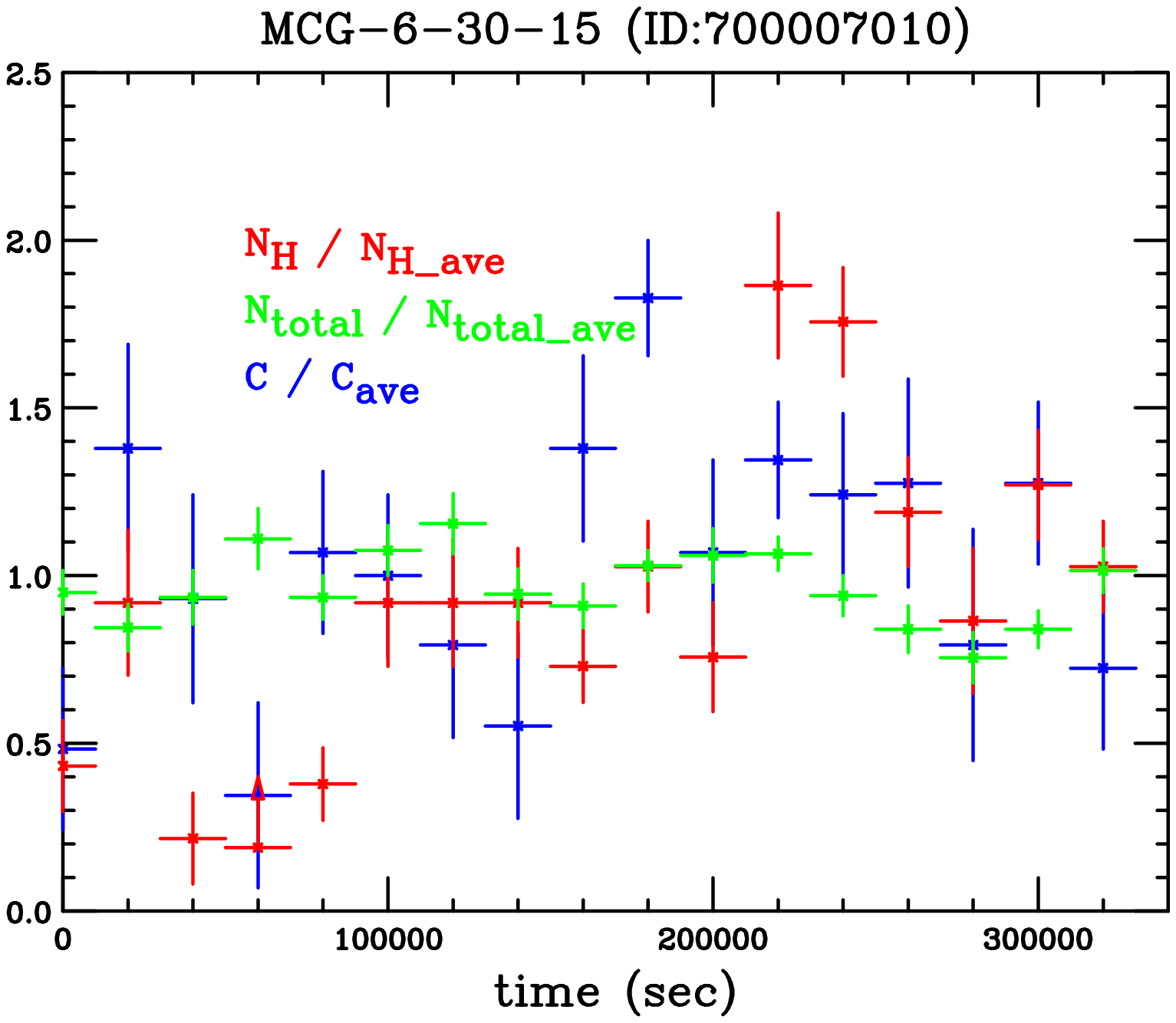}
\FigureFile(90mm,90mm){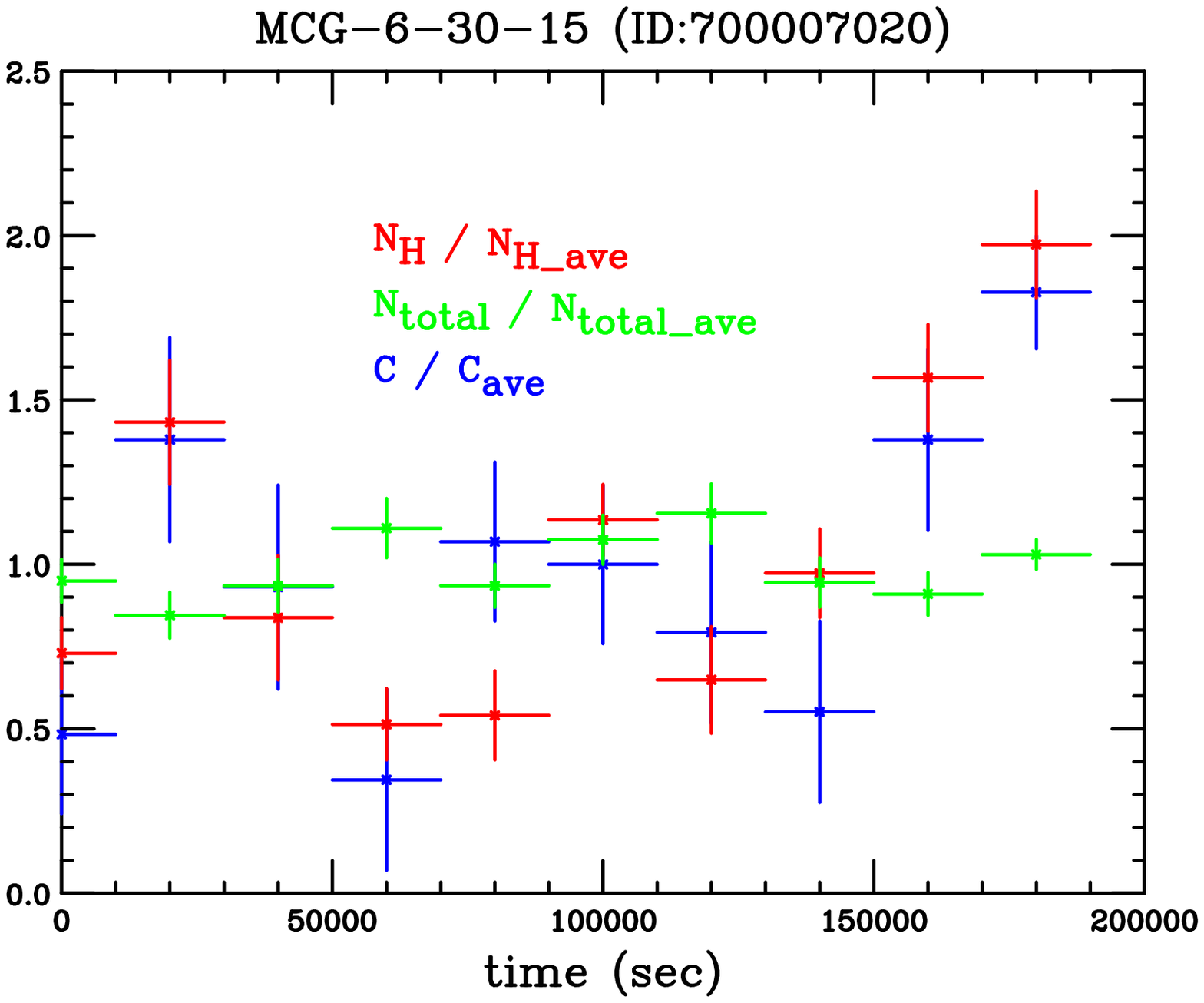}
\FigureFile(90mm,90mm){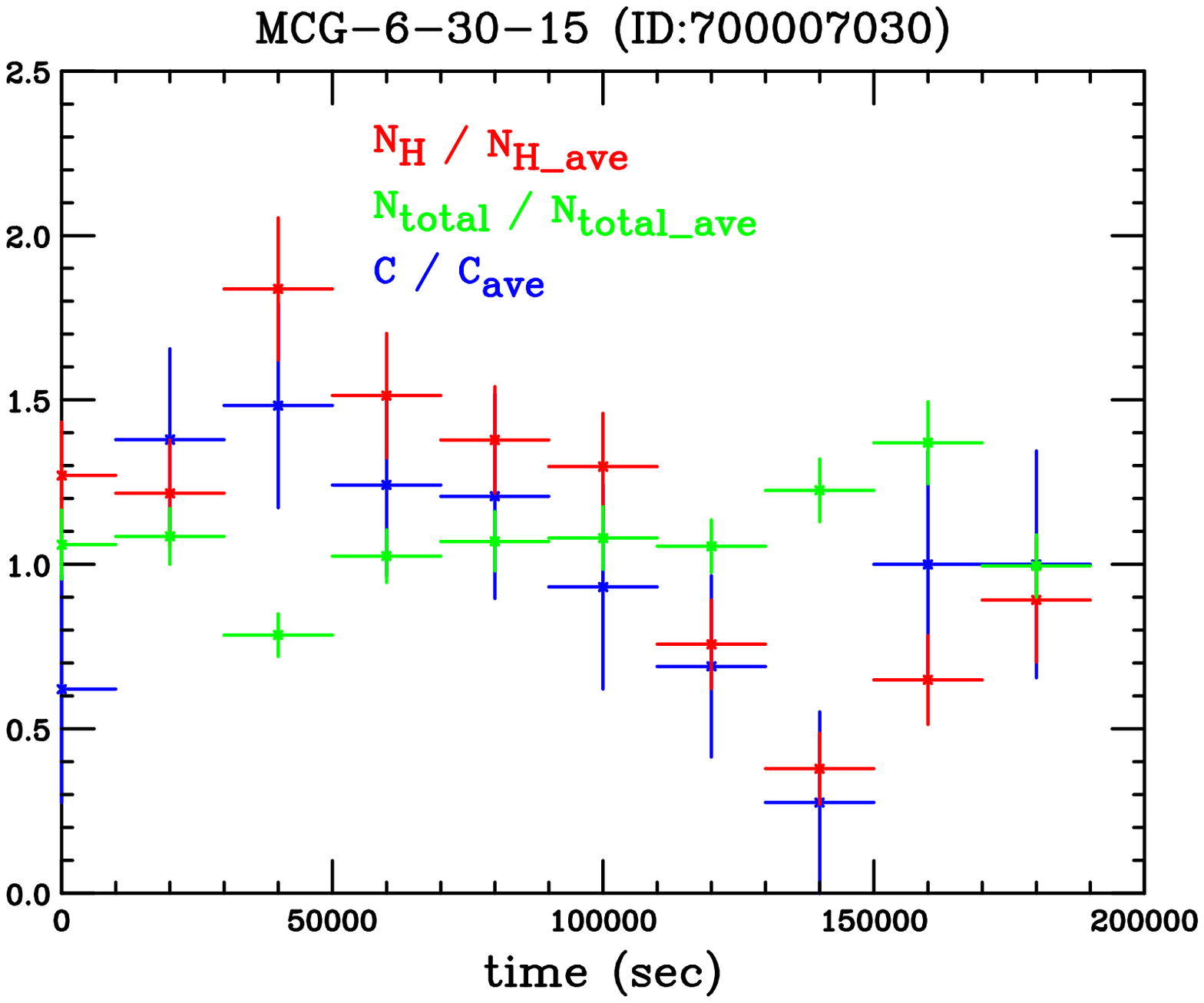}
\caption{Fractional variations of the total normalization, $N$, 
column density of the lower-ionized warm absorber, $N_{H, L}$, and the
covering fraction, $\alpha$,  for every 20 ksec.}
 \label{20ksec_par_var}
\end{figure}

 Obviously,
 \begin{equation}
  1- \alpha = \frac{ N_1 } { N }, 
 \end{equation}
 thus, instead of $(N_1,  N_2, N_{H,L})$, we can equivalently  use 
 $(N,  \alpha, N_{H,L})$ to describe the observed spectral variation.
 In Figure \ref{correlation_20ksec}, we show correlations
among  these three parameters.  

\begin{table}
\begin{center}
\caption{Variations of and correlations among $N$, $\alpha$ and $N_{H,L}$ in Figure \ref{correlation_20ksec}.}
\begin{tabular}{cccccc}\hline\hline
    &     &             & \multicolumn{3}{c}{Fitting results}\\\cline{4-6}
$x$ & $y$            & CC$^\dagger$ & $x=const. (\chi^2_r)$ & $y=const. (\chi^2_r)$ & $x = ay + b (\chi^2_r)$ \\ \hline
$N$ & $\alpha$       & $-0.30$     &   $x=0.0196 (2.35)$        & $y=0.31 (2.81)$         & $a=-0.003, b=0.02 (2.35)$\\
$N$ & $N_{H,L}$       & $-0.24$    &   $x=0.0196 (2.35)$        & N.A.                     & $a=-0.0001, b=0.02 (2.39)$\\
$\alpha$ & $N_{H, L}$ & $0.64$      &   $x=0.31 (2.81)$        & N.A.                      & $a=0.04, b=0.13 (1.63)$\\ \hline
\multicolumn{5}{l}{$^\dagger$Correlation Coefficient.}\\
\label{tab:correlation}
\end{tabular}
\end{center}
\end{table}

We study variations of and correlations among  the three parameters,  $(N,  \alpha, N_{H,L})$,  by calculating the correlation
coefficients between the two, as well as fitting the correlation with a constant parameter or a linear-function.
Results are summarized in Table \ref{tab:correlation}.
Consequently, we can see a good  ``orthogonality'' between 
 $N$ and $\alpha$, and  $N$ and $N_{H, L}$ (i.e., the fitting parameter $a$ is almost zero, where $N = a \alpha + b$ or $N = a N_{H,L} + b$). 
Namely, the total power-law
 normalization $N$ is not significantly variable, while the covering fraction,
 $\alpha$,  and  column-density of the low-ionized absorber,  $N_{H,L}$, are more variable. Furthermore, $\alpha$ and
$N_{H, L}$ indicate   a good  correlation with the correlation-coefficient
0.64.  Therefore, 
 spectral variation of MCG-6-30-15 at a timescale of 20 ksec is primarily
 described by  a single parameter, 
 either $\alpha$ or $N_{H,L}$.  We remind, however,  there is a hint of weak anti-correlation
between $N$ and $\alpha$.
Further sophistication of the model may explain this correlation (section \ref{sec:blocker}).
 
 Figure \ref{20ksec_par_var} indicates another representation of the spectral parameter variations.
  As a function of time, the three parameters,
 $N, \alpha, N_{H,L}$, respectively normalized by their average values, are indicated.
 The $\chi_r^2$ values
around the average (1.0)  are 2.4, 2.9 and 10.3 for $N/N_{ave}, 
\alpha/\alpha_{average}, N_{H,L}/N_{H,L,ave}$ respectively (d.o.f.=36).
 We can see $N$ is the least  variable, while $\alpha$ and $N_{H,L}$ 
 are in sync and  more significantly variable.
In the following, we will discuss physical meaning of these spectral 
variations.
 
 %




\section{Discussion}

 \subsection{Variable partial covering Model} \label{sec:vpcmodel}
\label{phenomenological_inte_section}
 We have seen that Suzaku and Chandra  energy spectra of 
 MCG--6-30-15  are successfully
described by the three-component model introduced with Equation 
(\ref{Eq2}).  
We found $N_2$, normalization of the heavily absorbed component,  is anti-correlated
with normalization of the direct component, $N_1$ (bottom panels
 of Figures \ref{relation_for_the_slice_spectra} and \ref{relation_for_the_slice_spectra_nH},
and Figure \ref{direct_absorbed_20ksec}).
The spectral variation 
indicates that the
``total normalization'', $N$,  (Equation \ref{Eq3}) is not so variable, 
while the covering fraction, $\alpha$,  (Equation \ref{Eq4}) and column-density of the
low-ionized warm absorber are in sync and more significantly variable
(Figure \ref{correlation_20ksec} and \ref{20ksec_par_var}).

 How should we interpret this characteristic spectral variation?
 %
Since  $N_1 $ and  $N_2 $ represent amounts of the 
non-absorbed and  absorbed fluxes emitted from the central X-ray source,  
respectively,  the anti-correlation  is  naturally 
 understood that  a rather constant X-ray source is partially obscured by absorbing 
matters with variable covering fraction, $\alpha$.
 The ``Difference Variation Function analysis'' introduced  by Inoue, 
Miyakawa and Ebisawa (2011) 
 also confirms that the ($N$, $\alpha$) parameter-set is more orthogonal and fundamental than 
 the ($N_1$, $N_2$) parameter-set.

Correlation between the low-ionized absorber $N_{H,L}$ and $\alpha$ 
(bottom panel of Figure \ref{correlation_20ksec}) is understood as follows:
 The low-ionized absorber represented by $W_L =  \exp -(\sigma(E, \xi) \: N_{H,L})$
  is optically thin (Table \ref{Miyakawa_model_ave_Suzaku}), 
 so that $W_L \approx 1 - \sigma(E, \xi) \: N_{H,L}.$
 On the other hand, the same absorption effect can be expressed 
 by a partial covering model  with a fixed column-density as,
 \begin{equation}
 W_L = 1 - \alpha + \alpha \exp (- \sigma(E, \xi) \: N_{H,L}^{(fixed)})\label{eq7}
 \end{equation}
 \begin{equation}
  \approx  1 -  \alpha\; \sigma(E, \xi) \: N_{H,L}^{(fixed)}.\label{eq8}
 \end{equation}
 Namely,  
 \begin{equation}
N_{H,L} = \alpha \; N_{H,L}^{(fixed)}, \label{eq9}
 \end{equation}
that is seen
in the bottom panel of Figure \ref{correlation_20ksec}.






 \subsection{Explanation of the spectral variation}
 
As we have seen, the  observed
 spectral variation of MCG-6-30-15 is primarily explained by
 variation of the partial covering fraction, $\alpha$.
Below, we designate our   model as  the ``variable partial covering  model''.
In the framework of the  variable partial covering  model,
the Suzaku intensity-sliced spectra in 1 -- 40 keV  is explained  mostly by 
change of the covering fraction.   
  Using the total normalization $N$ and the covering fraction $\alpha$, our three-component model
  (\ref{Eq2}) is rewritten as follows:
  \begin{equation}
  F = W_H W_L ( 1 - \alpha  + \alpha  W_2 ) N P + N_3 R P + I_{Fe}.\label{eq6}
  \end{equation}
In Equations (\ref{eq9}) and (\ref{eq6}),  only variable parameters are $\alpha$  and $N$.
In  Figure \ref{covering_fraction_change}
and  Table \ref{table1_slice}, we show an example of the
simultaneous fit of the  Suzaku intensity-sliced spectra with variable $\alpha$ (the same dataset as used in
section \ref{sec:sliced}).
The eight intensity-sliced spectra
are successfully fitted with  the constant normalization, $N = 1.80 \times 10^{-2}$ photons s$^{-1}$ cm$^{-2}$ at 1 keV
(except the brightest one where  $N$ is 1.5 times higher), 
while  $\alpha$ is variable from 0.63 (dimmest) to null (brightest).

 \subsection{Explanation of the  small variability in the iron energy band}

The well-known small variability in the iron energy band (Section 1) is also 
nicely explained in the framework of  the variable partial covering model.
In order to see energy
dependence of the variation at a specific timescale, we took the following
method to calculate the ``Difference  Variation Function'', $DVF(E)$ (Inoue,
Miyakawa and  Ebisawa  2011): (1) We choose a time-interval of   interest, $t$, to investigate  for spectral
variations; here, we
take $t=$40 ksec. (2)
Energy spectra are calculated for every $t/2$, namely, 20 ksec.  (3) For every two contiguous energy spectra, we recognize the ``brighter spectrum'' and the ``fainter  spectrum'', $B(E)$ and $F(E)$, respectively. (4) We average all the brighter spectra and all the fainter spectra.  
(5) Thus,
for a given time-period $t$, we create a single ``averaged bright spectrum''  $<B(E)>$ 
and a single ``averaged faint spectrum'' $<F(E)>$.
(6) We calculate difference of  the averaged bright spectrum and the averaged faint spectrum relative to the average, as
\begin{equation}
DVF(E) \equiv \frac{<B(E)> - <F(E)>}{<B(E)>+<F(E)>}.
\end{equation}

\begin{figure}[htbp]
\FigureFile(90mm,90mm){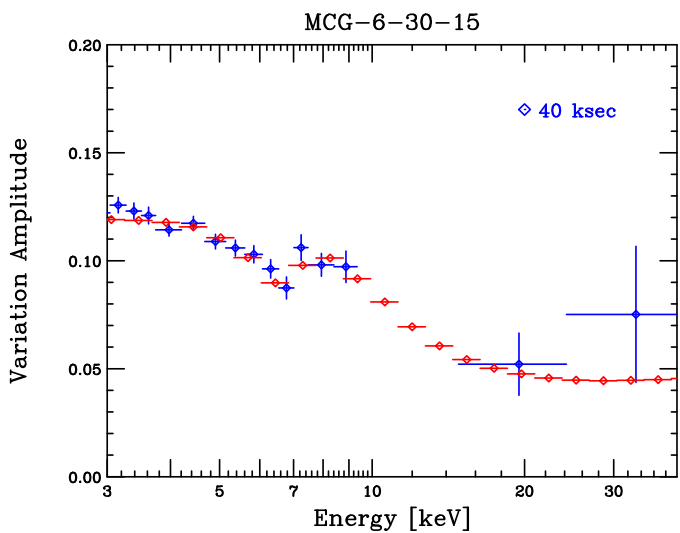}
\caption{Energy dependence of the  spectral variation  ("Difference Variation Function''; blue) and 
model simulation (red) in the framework of the variable partial covering model.
This figure is the same one as Figure 4 in Inoue, Miyakawa and Ebisawa (2011).
\label{cov_fac_simulation}
}
\end{figure}

In Figure \ref{cov_fac_simulation},  we show the Difference  Variation Function thus
calculated  at a timescale of 40 ksec.  
Together,  we show a model prediction of the Difference Variation Function
calculated in the framework of the variable partial covering model
(Eq. 29 in Inoue, Miyakawa and Ebisawa 2011), where  the covering fraction $\alpha$ is
assumed to be the major variable parameter, while the total normalization $N$ adds 
 minor fluctuation. Agreement of the observation and model calculation  is obvious.

\subsection{Parameters of   the warm  absorbers}
Next,  we estimate physical parameters of the warm absorbers in 
Equation (\ref{eq6}), $W_H$, $W_L$ and $W_2$, which are
high-ionized warm absorber, thin low-ionized warm absorber and  thick partial absorber,
respectively.  
We provide suffixes, $H, L$ and $2$ to distinguish column densities and ionization parameters 
of  these   warm absorbers.
In the following,    $r$ is distance from the
central X-ray source to the absorber, $h$ is a  representative  thickness of the absorbing
region along the line of sight, and
$n$ is density.

\begin{figure}[htbp]
\FigureFile(90mm,90mm){figure10.eps}
\caption{Suzaku intensity-sliced spectra simultaneously fitted with varying only the
covering fraction.  All the other parameters fixed, except the brightest spectrum 
 where the total normalization is 1.5 times greater than the others.
 \label{covering_fraction_change}}
\end{figure}

\setlength{\tabcolsep}{2.4pt}
\begin{table*}
\begin{center}
\scriptsize
\caption{Results of spectral fitting in 1--40 keV for the intensity-sliced spectra when only the covering fraction is varied (from slice 1 to 8,  the dimmest to  the brightest). }
\begin{tabular}{ccccccccc}
\hline
state  & slice 1 & slice 2 & slice 3 & slice 4 & slice 5 & slice 6 &
 slice 7 & slice 8  \\
\hline
\hline
\multicolumn{1}{l}{\em Interstellar absorption} \\
$N_H$ (${10}^{21}$ ${\rm cm}^{-2}$)  &  
		 \multicolumn{8}{c}{1.9$\pm$0.1} \\
\hline
\multicolumn{1}{l}{\em $W_H$} \\
$N_H$  (${10}^{23}$ ${\rm cm}^{-2}$)  &
		 \multicolumn{8}{c}{2.4 (fixed)} \\
log $\xi$  & \multicolumn{8}{c}{3.43$\pm$0.03} \\
\hline
\multicolumn{1}{l}{\em Covering Fraction} \\
$\alpha$   & 0.632 & 0.356 & 0.269 & 0.195 & 0.181 & 0.166 & 0.122 & $<0.1$\\
\hline
\multicolumn{1}{l}{\em $W_L$} \\
$N_{H,L}^{(fixed)}$  ($10^{22}\;{\rm cm}^{-2}$)  & \multicolumn{8}{c}{1.3} \\
$N_H$   & \multicolumn{8}{c}{$\alpha \;N_{H,L}^{(fixed)}$} \\
log $\xi$  & \multicolumn{8}{c}{1.88$\pm$0.03} \\
\hline
\multicolumn{1}{l}{$N$}\\
10$^{-2}$ph/s/{\rm cm}$^2$  at 1 keV & \multicolumn{7}{c}{1.80}  & 2.71   \\
\hline
\multicolumn{1}{l}{$N_1$}\\
 &  \multicolumn{8}{c}{$(1 - \alpha)\; N$}   \\
\hline
\multicolumn{1}{l}{$N_2$}\\
 &  \multicolumn{8}{c}{$ \alpha\; N$}   \\
\hline
\multicolumn{1}{l}{$W_2$}\\
$N_H$  (${10}^{24}$ ${\rm cm}^{-2}$)  &
		 \multicolumn{8}{c}{1.6 (fixed)} \\
log $\xi$  & \multicolumn{8}{c}{1.63$\pm$0.05} \\
\hline
\multicolumn{1}{l}{$P$}\\
photon index & 		 \multicolumn{8}{c}{1.88$\pm$0.01} \\
$E_{cut}$ (keV) & \multicolumn{8}{c}{160 (fixed)} \\
\hline
\multicolumn{1}{l}{$N_3$}\\
pexrav K (10$^{-3}$)  & 
		 \multicolumn{8}{c}{ 0.3 $\times$ $N_1$ (for  slice5) } \\
\hline
\multicolumn{1}{l}{$R^\dagger$}\\
cosIncl & \multicolumn{8}{c}{0.866 (fixed)} \\
\hline
\multicolumn{1}{l}{$I_{Fe}$}\\
line E (keV) & \multicolumn{8}{c}{6.35 (fixed)} \\
sigma (keV) &  \multicolumn{8}{c}{0.01 (fixed)} \\
norm (${10}^{-5}$ph/s/cm$^2$) & \multicolumn{8}{c}{1.4$\pm$0.2} \\
\hline
edge E (keV) & \multicolumn{8}{c}{7.11 (fixed)} \\ 
MaxTau & \multicolumn{8}{c}{0.08$\pm$0.01} \\ 
\hline
line E (keV) &  \multicolumn{8}{c}{7.0 (fixed)} \\
sigma (keV) & \multicolumn{8}{c}{0.01 (fixed)} \\
norm (${10}^{-6}$ph/s/cm$^2$) & \multicolumn{8}{c}{$-4.8\pm$1.7} \\
\hline
line E (keV) &  \multicolumn{8}{c}{2.35$\pm$0.01} \\
sigma (keV) & \multicolumn{8}{c}{0.01 (fixed)} \\
norm (${10}^{-5}$ph/s/cm$^2$) & \multicolumn{8}{c}{$-1.9\pm$0.3} \\
\hline
Reduced chisq (d.o.f) & \multicolumn{8}{c}{0.49 (1149)} \\
\hline\hline
\multicolumn{2}{l}{$\dagger$ ``pexrav'' model in xspec is used.}\\
\label{table1_slice}
\end{tabular}
\end{center}
\end{table*}

First, in general, we note the following relations hold:
\begin{equation}
\xi \equiv {L}/{nr^2} = \frac{L}{N_H r} \left(\frac{h}{r}\right), \; {\rm where} \; N_H = nh,
\end{equation}
\begin{equation}
r = \frac{L}{ {N_{H}} {\xi}} \left(\frac{h}{r}\right) \leq \frac{L}{{N_{H}} {\xi}}. 
\end{equation}

For the luminosity, we take the representative value $10^{43}$ erg/s,
which is valid  for order estimation, since the   flux  variation is much less than an 
order of magnitude.


\subsubsection{High-ionized warm absorber, $W_H$} 

From model fitting (Table \ref{table1_slice}), ${\xi}_H$ $\simeq$ 10$^{3.4}$ erg cm/s, $N_{H,H}$ $\simeq$ 10$^{23.4}$ cm$^{-2}$.  Hence, 
\begin{equation}
r_H = \frac{L}{ {N_{H,H}} {\xi}_H} \left(\frac{h_H}{r_H}\right) \leq \frac{L}{{N_{H,H}} {\xi}_H} \simeq 10^{16.2} \: {\rm cm}.
\end{equation}

Our analysis has
 indicated that the high-ionized warm absorber is constant, while partial covering fraction
is variable.  This suggests the variable partial absorbers be embedded in the largely extended, 
static high-ionized warm absorber.  Hence, we  assume $h_H$ $\sim$ $r_H$, and take $r_H \sim$ 10$^{16}$ cm and  $n_H \sim$ 10$^{7}$ cm$^{-3}$ for typical parameters of the high-ionized warm absorbers.

\subsubsection{Thick partial absorber, $W_2$}

From model fitting (Table \ref{table1_slice}),  ${\xi}_2 \simeq$ 10$^{1.6}$ erg cm/s, 
$N_{H,2} \simeq$ 10$^{24.2}$ cm$^{-2}$. 
Hence, 
\begin{equation}
r_2 \simeq \frac{L}{ {N_{H,2}} {\xi}_2} \left(\frac{h_2}{r_2}\right) \leq \frac{L}{
{N_{H,2}} {\xi}_2} \simeq 10^{17.2} \; {\rm cm}.
\end{equation}
Here, we may further constrain the parameters of the
partial absorber from variation timescale 
of the partially absorbed component.   In our variable partial covering model, typical  variation
timescale ($\sim10^5$ sec; Figure \ref{20ksec_par_var})  corresponds to the crossing
time of the   X-ray absorbing clouds  in front of the X-ray source.
We will call  these clouds   as the
``low-ionized clouds'', hereafter. 
Assuming that  size of the X-ray source and that of a  low-ionized cloud
have similar dimensions, the crossing timescale
is expressed as $\sim h_2/V$, where $V$ is the velocity of a low-ionized cloud.
Hence, 
\begin{equation}
h_2 \sim V \times 10^{5} \sim 10^{14} \left(\frac{V}{10^{9} \; {\rm cm/s}}\right) \; {\rm  cm}, \label{sizeofcloud}
\end{equation}
where $V$ is normalized to a typical velocity of the broad line region (BLR) clouds.
Combining $h_2$ and $N_{H,2} \approx 10^{24}\; {\rm cm}^{-2}$, we estimate parameters of 
the partial absorbers as follows:
\begin{equation}
r_2 \simeq \left(\frac{L}{{\xi}_2 N_{H,2}} h_2 \right)^{1/2} \simeq 10^{15.7} \; {\rm cm},   \label{eq:r2}
\end{equation}
\begin{equation}
n_2 \sim 10^{10}\; {\rm cm}^{-3}. \label{n_2}
\end{equation}
Note that a low-ionized cloud may have internal ionization structures, which we will see below.

\subsubsection{Thin low-ionized warm absorber, $W_L$ }

In our variable partial covering model, 
the low-ionized warm absorber ($W_L$)
and the partial absorber ($W_2$) are closely associated.  
 Presumably, they are 
 different parts of the  low-ionized clouds.
We point out that presence of such internal structures
in the BLR clouds is also suggested  in Mrk 766 (Risaliti et al.\ 2011).

In  our  model,  the low-ionized warm absorber has a 
fixed column-density  $N_{H,L}^{(fixed)}$ with
a variable partial covering factor
(Equation \ref{eq9});  we simply write this column density as $N_{H,L}$ below.
From model fitting (Table \ref{table1_slice}), 
$N_{H,L} \sim 10^{22.1}$ cm$^{-2}$ and ${\xi}_L \simeq$ 10$^{1.9}$ erg cm/s. 


If we take the distance of the low-ionized clouds $r_2$, 
\begin{equation}
n_L =\frac{L}{\xi_L r_2^2} \simeq 10^{9.7} \; {\rm cm^{-3}}, 
\end{equation}
\begin{equation}
h_L =N_{H,L}/n_L \simeq 10^{12.4} \; {\rm cm}.
\end{equation}
Note that values of $n_L$ and $\xi_L$ are  between those 
of the high-ionized absorber ($n_H \sim 10^7$ cm$^{-3}$ and $\xi_H\sim 10^{3.4}$) 
and the partial absorber ($n_2 \sim 10^{10}$ cm$^{-2}$ and $\xi_2 \sim 10^{1.6}$),
respectively. The thickness $h_L$ is much smaller
than that of the partial absorber $h_2 \sim 10^{14}$ cm.
Therefore, the low-ionized absorber is considered to be
 in the boundary layer  between the low-ionized cloud and the
high-ionized warm absorber.  We may call this putative layer as the ``cloud envelope''.

Meanwhile, it is known that there takes place  thermal
instability in the range of $\xi =  10^{2}\sim 10^{3}$ erg cm/s (Reynolds $\&$ Fabian 1995).  
Therefore, it is reasonable that 
the cloud envelope continuously connects the  partial absorber in the
low-ionized cloud 
and the surrounding constant high-ionized warm absorber.



\subsubsection{Putative X-ray blocker}\label{sec:blocker}

We notice the Thomson optical depth ($\sim N_H/10^{24.2}$) of the partial absorber is about unity.
Presumably,  this is not just a coincidence, but suggests that the low-ionized 
cloud has internal  density gradients extending up to more than unit Thomson optical depth, 
such that attenuation with 
continuously different optical depths is approximated by  unit Thomson optical depth.
Thus, the low-ionized cloud 
may well have  central, cold  Thomson thick cores ($N_H \geq 10^{24.2}$ cm$^{-2}$), 
which are completely opaque to incoming X-rays.
We may not see these ``X-ray blockers'' directly, but if we take into account the effect of
the X-ray blockers, the weak anti-correlation between the total normalization $N$ and the
covering fraction $\alpha$  (top-panels of Figure \ref{correlation_20ksec}) might be explained.

\begin{table*}\begin{center}
\caption{Summary of Parameters of the Absorbers}
\begin{tabular}{lccccccc}\hline\hline
\multicolumn{3}{c}{Absorbers}& $r$ [cm]  & $\log N_H$  & $\log \xi$ & $h$ [cm]   & $n$ [cm$^{-3}$]\\\hline
\multicolumn{3}{l}{High-ionized warm absorber, $W_H$}& $10^{16}$  & 23.4        & 3.4        &  $10^{16}$ & $10^7$ \\
Low-ionized cloud &$\left\{ 
                  \begin{tabular}{c}
                  Envelope\\
                  Main body\\
                  Core  \\
                  \end{tabular}\right.$
                  &$\left.
                   \begin{tabular}{c}
                  Thin low-ionized warm absorber, $W_L$\\
                  Thick partial absorber, $W_2$\\
                  X-ray Blocker  \\
                  \end{tabular}\right\}$
                                                      &$10^{15.7}$&$\left\{ 
                                                                  \begin{tabular}{c}
                                                                  22.1\\
                                                                  24.2\\
                                                                  $\geq24.2$  \\
                                                                  \end{tabular}\right.$
                                                                &$\left. 
                                                                  \begin{tabular}{c}
                                                                  1.9\\
                                                                  1.6\\
                                                                  $\leq 1.6$  \\
                                                                  \end{tabular}\right.$
                                                                &$\left. 
                                                                  \begin{tabular}{c}
                                                                  $10^{12.4}$\\
                                                                  $10^{14}$\\
                                                                  $\approx 10^{14}$\\
                                                                  \end{tabular}\right.$
                                                                &$\left. 
                                                                  \begin{tabular}{c}
                                                                  $10^{9.7}$\\
                                                                  $10^{10}$\\
                                                                  $\geq 10^{10}$\\
                                                                  \end{tabular}\right\}$\\ \hline
\label{tab:absorbers}
\end{tabular}
\end{center}\end{table*}

\subsection{Origin of the  Variable Partial Covering}
Finally, we present a physical picture surrounding the X-ray source 
to explain the observations.  
In Table \ref{tab:absorbers}, we summarize parameters of the absorbers in our model.
Basic points of our variable partial covering model are the following:
\begin{enumerate}
\item Observed X-ray flux and spectral variability is primarily caused by 
 random passages of the
 low-ionized clouds in front of the central X-ray source. A
 typical size of  each low-ionized cloud is $h_2 \sim 10^{14}$ cm 
and a typical velocity  is $V \sim10^9$ cm/s (Equation \ref{sizeofcloud}).
\item Central X-ray source is not always fully blocked,  nor  
 fully exposed.
The covering fraction is variable between null to $\sim0.6$
(Table \ref{table1_slice}), and it fluctuates by 
an amplitude of $\sim50$ \% (Figure \ref{20ksec_par_var}).    
\end{enumerate}

We may estimate the average covering fraction $C_{ave}$ from intensity-sliced
 spectra as 
 $C_{ave}$ = $\displaystyle  {\sum}^{8}_{i=1} C_i \times t_i
/T$, where $t_i$ and $T$ are exposure time for each spectrum
and the total observation,  respectively. 
We thereby obtain $C_{ave}
 \sim$ 0.3. 
The integrated covering fraction is considered to be about 0.1 for BLR
(e.g.,  Blandford et al.\ 1991). 
Thus, our results suggest that the partial absorption takes place in the  BLR,
and the low-ionized clouds corresponds to the BLR blobs.

We may estimate number of the absorbing clouds in the field of view, $N_{cloud}$,
and size of the central X-ray source,  $r_x$,     as follows.
Assuming that fluctuation of the covering fraction, $\approx 0.5$, 
is determined by statistical fluctuation of the 
number of clouds, $\Delta N_{cloud}/N_{cloud} $, 
 $N_{cloud}$ is estimated as $\approx4$.
The average covering fraction, $\approx0.3$,  may be written as  
$N_{cloud}  \pi ({h_2/2})^2/\pi {r_x}^2$, thus, 
$r_x \approx 2 h_2 \approx 2 \times 10^{14}$ cm.

Typically, BLR blobs have a velocity of  $V$  $\sim$
10$^{9}$ cm/sec.
Assuming the  Kepler velocity has a similar value, 
 $r = (c^2/2 V^2) r_s \sim$ 500 $r_s$, where
$r_s$ is the Schwarzschild radius.
Identifying this with  $r_2 \sim$ 10$^{15.7}$ cm (Equation \ref{eq:r2}), 
the central black hole mass of MCG-6-30-15 is $\sim 3 \times
10^{7} M_{\odot}$.  Thus, the central X-ray
source ($r_x \sim 2 \times 10^{14}$ cm)  is extended to  $\sim 20 \; r_s$
around the black hole.  Besides variation of the BLR coluds,
the  X-ray source is variable on a timescale down to  $\sim$
1 ksec (Inoue, Miyakawa and Ebisawa 2011), 
which may be compared with the
free fall time,   $r/\sqrt{\frac{GM}{r}}$. 
Then, the intrinsic source variation  arises  at a few times the 
Schwarzshild radius, around the  innermost  stable circular orbit.

We point out that solid angle of the low-ionized clouds seen from the X-ray source 
should be  much smaller than 4$\pi$,
otherwise we will have to see strong fluorescent lines from the clouds.
If a thick  absorber with $N_H \gtrsim 10^{24}$ cm$^{-2}$ covers the power-law component
completely ($\Omega \sim 4 \pi$), equivalent width of the iron line to the 
reflection component is $\sim$ 1 keV. 
Considering   $C_{ave} \sim$ 0.3 and 
the ratio of the  reflection component to the direct component, 10 to 1,  the
expected iron line  equivalent
width is $\sim$ 30 eV. 
As mentioned in subsection 5.5.2, we found the fit improves if a 
mildly broad disk-line is added,
where the best-fit  inner radius and the  equivalent width 
are $\gtrsim$ 200 $r_g$ and $40\pm20$ eV, respectively (Table \ref{Miyakawa_diskline_model_ave_Suzaku}). 
These mildly broadened,  weak iron emission lines
may well be expected  from fluorescence in the  BLR clouds.

\begin{figure*}[htbp]
\begin{center}
\FigureFile(140mm,140mm){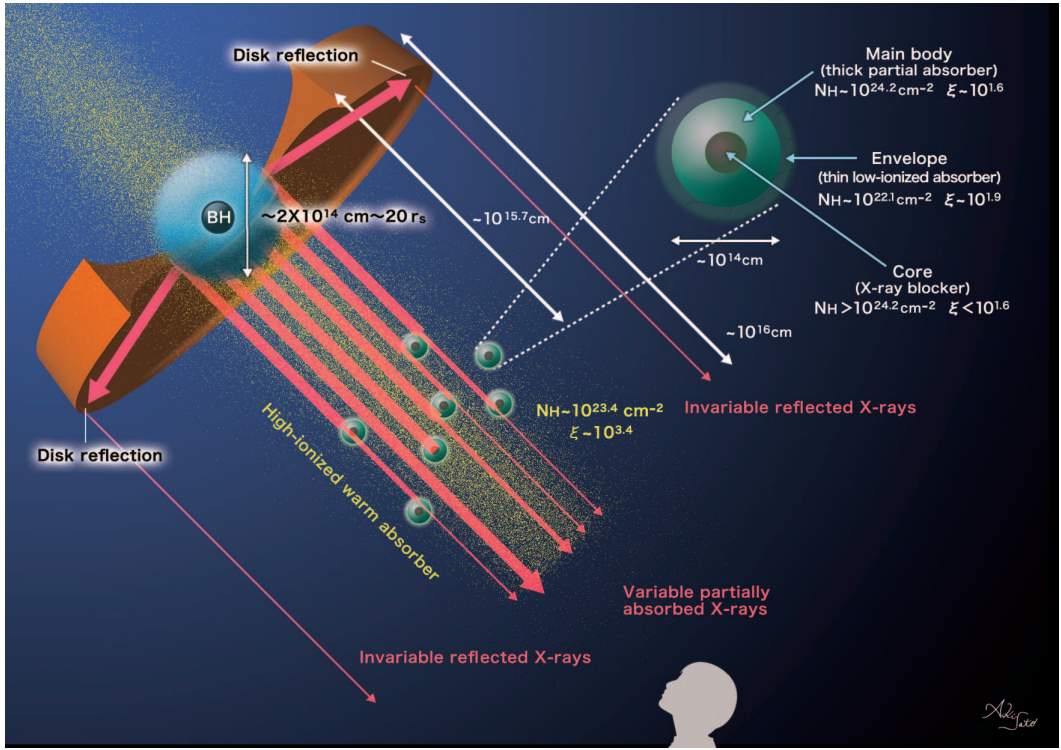}
\end{center}
\caption{Schematic picture of the variable partial covering  model 
for MCG-6-30-15 and internal 
structure of the absorbing clouds.}
 \label{picture_of_region}
\end{figure*}

Figure $\ref{picture_of_region}$ gives a schematic view of 
internal structure of the low-ionized cloud and  the variable partial
covering model surrounding the black hole.
The low-ionized cloud (BLR cloud) has internal ionization structure.
A putative  blocker  in the core would block the
incoming X-rays completely.
 The main body of the low-ionized cloud is the thick partial absorber surrounding
 the core. Still the outer cloud envelope corresponds to the thin
low-ionized warm absorber.  Central X-ray emission region around the
black hole is mildly extended, and 
 partially covered by these low-ionized clouds. Variation of the covering
fraction explains most of the observed flux and spectral variations.

Finally we comment that the idea of partial covering 
is not new for MCG-6-30-15 to explain its spectral shape and varaition
(e.g., Matsuoka et al.\ 1990; McKernan and Yaqoob 1998;
Miller, Turner and Reeves 2008, 2009).  Recently, similar partial covering
models are proposed to explain spectral variations of 
NGC 3516 (Turner et al.\ 2008) ,  Mrk 766  (Risaliti et al.\ 2011)
and other Seyfert galaxies (Turner \& Miller 2009).
Presumably, partial covering of rather constant X-ray source is a common mechanism in Seyfert galaxies.
It is of interest to study spectral variations of Seyfert galaxies systematically to see
if the observed X-ray variablity is truely due to intrinsic luminosity
variation or explained by partial covering of the constant source.

\section{Conclusion}

We have analyzed Suzaku and Chandra archival data of MCG-6-30-15 
and constructed a new model to explain the observed spectral variation.
 Our main conclusions are summarized below:

\begin{enumerate}
\item We have shown that  the observed energy spectra
and spectral variation can be explained by  the ``three-component model'' 
(Equation \ref{Eq2}) that includes (1) a direct power-law component, (2) 
a heavily absorbed power-law component by thich photoionized material, 
and (3) a cold disk reflection component far from the black hole.
The first two components are affected by two warm abosrbers having different ionization states.

\item  The ionized iron K-edge of the heavily
	  absorbed component and  spectral curvature due to the warm
	  absorbers  explain most of  the seemingly  broad ``disk line'' spectral
	  feature.  Consequently,
  general relativistic interpretation of the 
	   broad iron emission line feature in MCG-6-30-15 is not confirmed.
The claim that
	   MCG-6-30-15 is a Kerr black hole with nearly extreme rotation assuming the
relativistic line-broadening (e.g.,
	   Miniutti et al. 2007) is questionable.

\item  We propose the ``variable partial covering 
 	   model'', in which a  central, moderately extended 
 X-ray source is partially covered by   variable low-ionized clouds  
in the  line of sight.
In this model, observed X-ray spectral variation 
is primarily caused  by  change of the partial 
 covering fraction of the central source.
These  absorbing  clouds  presumably correspond to the fast-moving broad line region 
(BLR) clouds.
\end{enumerate}

The authors would like to thank Profs.\  P. \.Zycki, Y. Terashima, Y. Ueda and K. Makishima for
fruitful discussion.
This research has made use of public Suzaku data obtained through the Data ARchives
and Transmission System (DARTS), provided by the  
Institute of Space and Astronautical Science (ISAS) at Japan
Aerospace Exploration Agency (JAXA), public Chandra data obtained through the High Energy Astrophysics Science Archive Research Center (HEASARC) at NASA/Goddard Space Flight Center.

For data reduction, we used the software provided by the
High Energy Astrophysics
Science Archive Research Center (HEASARC) at 
NASA/Goddard Space Flight Center
and Chandara X-ray Center (CXC) at Harvard-Smithsoniann Center for Astrophysics.
K.E. acknowledges Ms.\ Akiko Sato for helping us illustrate our model
 and producing a beautiful artwork.

\end{document}